\documentclass[a4paper,11pt]{article}
\pdfoutput=1 

\usepackage{jinstpub} 
\usepackage{siunitx}
\usepackage{graphicx}
\usepackage{subfig}
\usepackage{appendix}
\usepackage[nopostdot,nogroupskip,style=super,nonumberlist,toc,automake,toc]{glossaries} 

\makeglossaries

\newglossaryentry{TiPC}{name={TiPC},description={Timed Photon Counter}}
\newglossaryentry{PMT}{name={PMT},description={photomultiplier tube}}
\newglossaryentry{SE}{name={SE},description={secondary electron}}
\newglossaryentry{PE}{name={PE},description={primary electron}}
\newglossaryentry{TSE}{name={TSE},description={transmission secondary electron}}
\newglossaryentry{FSE}{name={FSE},description={forward-scattered electron}}
\newglossaryentry{RSE}{name={RSE},description={reflection secondary electron}}
\newglossaryentry{BSE}{name={BSE},description={backscattered electron}}
\newglossaryentry{TEY}{name={TEY},description={(total) transmission electron yield}}
\newglossaryentry{ALD}{name={ALD},description={atomic layer deposition}}
\newglossaryentry{DRIE}{name={DRIE},description={deep-reactive ion etching}}
\newglossaryentry{PR}{name={PR},description={photoresist}}
\newglossaryentry{PECVD}{name={PECVD},description={plasma-enhanced vapor deposition}}
\newglossaryentry{SEM}{name={SEM},description={scanning electron microscope}}
\newglossaryentry{tynode}{name={tynode},description={transmission dynode}}

\title{\boldmath Ultra-thin corrugated metamaterial film as large-area transmission dynode}

\author[a,b,1]{H.W. Chan\note{Corresponding author.}}
\author[a,b]{V. Prodanovi\'c}
\author[c]{A.M.M.G. Theulings}
\author[b]{T. ten Bruggencate}
\author[c]{C.W. Hagen}
\author[b]{P.M. Sarro}
\author[a]{and H. v.d Graaf}

\affiliation[a]{National Institute for Subatomic Physics (NIKHEF),\\Science Park 105, 1098 XG, Amsterdam, The Netherlands}
\affiliation[b]{Faculty of Electrical Engineering, Mathematics, and Computer science, Department of microelectronics/ECTM,\\Feldmannweg 17, 2628 CT, Delft, The Netherlands}
\affiliation[c]{Faculty of applied sciences, Department of Imaging Physics, Delft University of Technology,\\Lorentzweg 1. 2628 CJ, Delft, The Netherlands}

\emailAdd{h.w.chan@hotmail.com}

\abstract{Large-area transmission dynodes were fabricated by depositing an ultra-thin continuous film on a silicon wafer with a 3-dimensional pattern. After removing the silicon, a corrugated membrane with enhanced mechanical properties was formed. Mechanical metamaterials, such as this corrugated membrane, are engineered to improve its strength and robustness, which allows it to span a larger surface in comparison to flat membranes while the film thickness remains constant. The ultra-thin film consists of three layers (Al\textsubscript{2}O\textsubscript{3}/TiN/Al\textsubscript{2}O\textsubscript{3}) and is deposited by atomic layer deposition (\gls{ALD}). The encapsulated TiN layer provides in-plane conductivity, which is needed to sustain secondary electron emission. Two types of corrugated membranes were fabricated: a hexagonal honeycomb and an octagonal pattern. The latter was designed to match the square pitch of a CMOS pixel chip. The transmission secondary electron yield was determined with a collector-based method using a scanning electron microscope. The highest transmission electron yield was measured on a membrane with an octagonal pattern. A yield of 2.15 was achieved for \SI{3.15}{\keV} incident electrons for an Al\textsubscript{2}O\textsubscript{3}/TiN/Al\textsubscript{2}O\textsubscript{3} tri-layer film with layer thicknesses of 10/5/\SI{15}{\nm}. The variation in yield across the surface of the corrugated membrane was determined by constructing a yield map. The active surface for transmission secondary electron emission is near 100\%, i.e. a primary electron generates transmission secondary electrons regardless of the point of impact on the corrugated membrane.}


\arxivnumber{2008.09054} 

\begin{document}
\maketitle
\flushbottom

\printglossary[title={List of Abbreviations}] 


\section{Introduction}
\label{sec:intro}

\begin{figure}[b!]
\begin{minipage}[t]{0.45\linewidth}
\includegraphics[width=\linewidth,origin=c,angle=0]{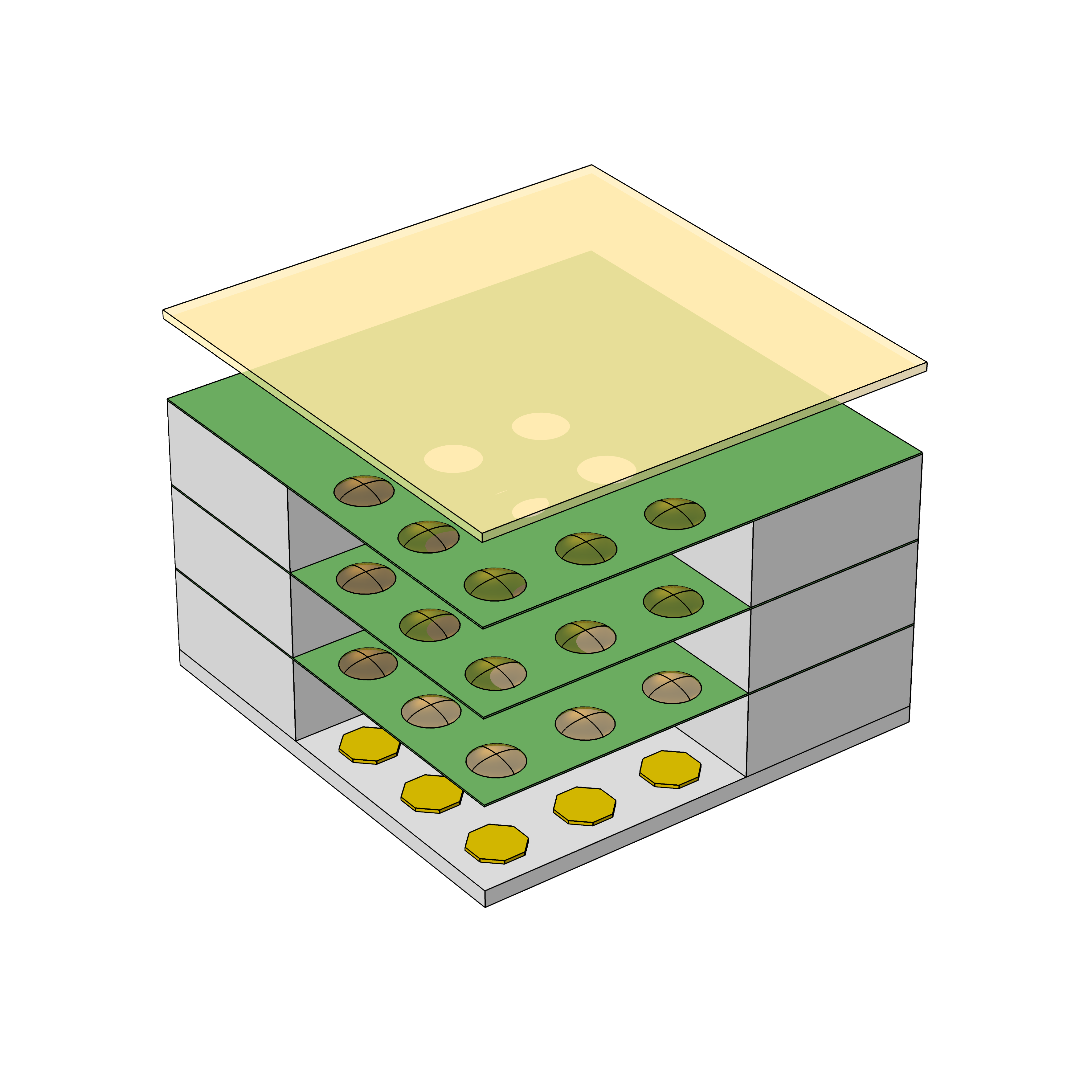}
\caption{\label{fig:1} Schematic drawing of TiPC with tynodes comprising ultra-thin circular membranes suspended in a supporting mesh. A photon is converted to a photoelectron at the photocathode. Due to the electric field between each stage, the photoelectron will gain sufficient energy to induce transmission secondary electron emission at each tynode. After several multiplication stages, an avalanche of electrons will be collected by the pixel pads. The active membranes are suspended in small circular windows arranged in a 64-by-64 array with a pitch of \SI{55}{\um}, which matches the pixel pitch of a TimePix chip.}
\end{minipage}
\hfil 
\begin{minipage}[t]{0.45\linewidth}
\includegraphics[width=\linewidth,origin=c,angle=0]{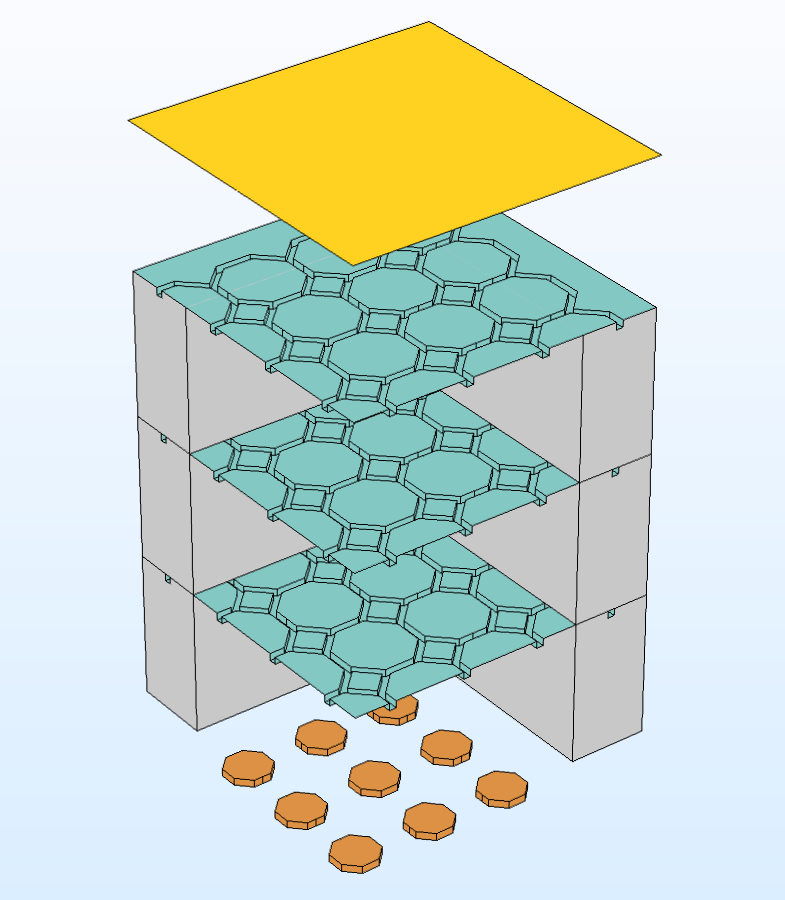}
\caption{\label{fig:2} Schematic drawing of TiPC with metamaterial tynodes. The metamaterial tynode consists of an ultra-thin corrugated membrane with uniform thickness. An incoming electron with sufficient energy can generate transmission secondary electrons regardless of the point of impact. However, the avalanche should be collected by the pixel pad in order for the photon/photoelectron to be detected. }
\end{minipage}
\end{figure}

The Timed Photon Counter (\gls{TiPC}) is a novel vacuum electron multiplier that has ultra-thin transmission dynodes (\gls{tynode}s) as multiplication stages \cite{VanderGraaf2017}. The detection principle is similar to a photomultiplier tube (\gls{PMT}), i.e. photoemission from a photocathode and subsequent (secondary) electron multiplication in vacuum. In a PMT, a photoelectron is multiplied within a series of (reflective) dynodes \cite{Hamamatsu2007}. The dynodes are carefully aligned in a sequence, where the electric potential is step-wisely increased between each dynode. Under influence of the electric field, the secondary electrons (\gls{SE}s) are accelerated and directed from dynode to dynode. On each impact, their number is multiplied. Eventually, the avalanche of electrons is collected by the anode, which provides an output signal. 

PMTs are widely used as photodetector in low-light applications due to its high gain, low noise and large sensing area. Their spectral response can be tailored to a certain part of the spectrum, ranging from ultraviolet to near-infrared, by choosing a window material with high transparency and a photocathode material with high sensitivity to the selected region \cite{Hamamatsu2007}. However, the design of PMTs has some drawbacks due to the complex arrangement of the (reflective) dynodes. First, the non-uniform electron paths between each dynode limit the time resolution in the order of nanoseconds. Second, the electron paths are easily perturbed by magnetic fields, which precludes PMTs to be used in applications with strong magnets. Last, the device is voluminous, fragile and expensive to manufacture. 

TiPC will improve upon PMTs in terms of time and spatial resolution. The core innovation is the use of tynodes as multiplication stages and a CMOS pixel chip as read-out. A tynode is an ultra-thin membrane that is optimized for transmission secondary electron emission. An incoming primary electron (\gls{PE}) will generate multiple transmission secondary electrons (\gls{TSE}s) within the membrane as the PE transmits through it. This allows tynodes to be closely stacked on top of each other. This design/configuration has many benefits. First, the electric fields between the planar tynodes are more uniform and stronger. The electron paths are more uniform, which reduces the electron transit time spread and improves the overall time resolution. Second, the increased electric field strength makes TSEs less susceptible to external magnetic fields. Third, 2D spatial information can be acquired using a CMOS pixel chip as read out. Last, TiPC is a compact planar photodetector.

The secondary electron emission process can be divided in three steps: \emph{generation}, \emph{transport} and \emph{escape} of internal SEs \cite{Bruining1954}. The first step describes the interaction of PEs within matter. PEs that enter matter will scatter and lose energy. Some of the energy is used to generate internal SEs. The second step models the transport of these internal SEs. The distance that an internal SE can travel depends on the material. Often the band gap model is used to describe the difference in transport in metal, semiconductors and dielectrics. In general, wide band gap materials allow internal SEs to travel a greater distance before they can either escape or be reabsorbed. The third step describes the escape process: internal SEs that reach the surface will encounter a barrier and need sufficient energy to cross it. The amount of energy that is needed is determined by the work function (metals) or the electron affinity (semiconductors and dielectrics) of the material. 

Thus, the amplification of a tynode depends on the material of the tynode and its surface condition. The performance of a tynode is characterized by the maximum transmission electron yield $\sigma_T^{\text{max}}$ obtained with PEs with energy $E_T^\text{max}$; or in a more compact notation: $\sigma_T^{\text{max}}(E_T^\text{max})$. In a detector with multiple tynodes, the maximum transmission electron yield determines the overall gain, while the required PE energy determines the operating voltage of the detector. The PE energy $E_T^\text{max}$ is determined by the film thickness and the range of the electron in the film. Only SEs that are generated near the exit surface of the film have a chance to escape. Therefore, the film thickness is a parameter that needs to be optimized. In the past, several types of transmission dynodes with high transmission electron yields (\gls{TEY}) have been reported \cite{Tao2016}. Unfortunately, the maximum TEY was only achieved with PEs with very high energy. For instance, a TEY of $27(\SI{9}{\keV})$ was reported for a caesium-activated CsI film deposited on an Al/Al\textsubscript{2}O\textsubscript{3} film by Hagino et al.\cite{Hagino1972}. 

Hence, we used Micro-Electro-Mechanical System technology to fabricate ultra-thin membranes, which were suspended within a supporting mesh with circular windows with a diameter of $\SI{30}{\um}$ \cite{Prodanovic2018}. In figure \ref{fig:1}, a schematic drawing of TiPC with these tynodes is shown. The windows are arranged in a 64x64 array with a pitch of $\SI{55}{\micro\meter}$ to match the spacing of the pixel pads on a TimePix chip \cite{Llopart2007}. An aligned tynode stack will form 'channels' in which electron multiplication will take place and be read out by the individual pixels. Two types of material were used to form the ultra-thin membrane: Low-pressure-chemical-vapor-deposition silicon nitride and atomic-layer-deposition (ALD) aluminum oxide. The former has a TEY of 1.6 ($\SI{2.85}{\keV}$) for a membrane with a thickness $d=\SI{40}{\nm}$, while the latter has a TEY of 2.6 ($\SI{1.45}{\keV}$) for a membrane with $d=\SI{10}{\nm}$. On both tynodes, a titanium nitride (TiN) layer was deposited on the PE entrance side to provide in-plane conductivity. In a different process, the TiN layer was encapsulated within two layers of Al\textsubscript{2}O\textsubscript{3} to improve the reliability of the fabrication process \cite{Chan2020}. This tri-layer film had a TEY of 3.1 (1.55 keV). The film layer was deposited on a flat substrate.

The supporting mesh is a necessity in the design of the tynode array due to the fragility of ultra-thin films. However, there are some drawbacks as well. First of all, it reduces the sensing area of the detector: photoelectrons that land on the supporting mesh will not be detected. The collection efficiency of TiPC can be estimated by using the ratio of the surface area of the windows and the mesh. For an array of windows with a diameter of $\SI{30}{\um}$ and a pitch of $\SI{55}{\um}$, the active surface area is only $23.4\%$. Second, the collection efficiency also depends on the alignment of the windows. Misalignment can potentially result in loss of SEs in the tynode stack. Last, SEs can be trapped within the dielectric material (silicon nitride) of the mesh. During prolonged irradiation, charge can accumulate on the exposed surface of the dielectric mesh and cause distortions in the electric field.

Therefore, new tynodes have been made with some improvements, such as increased window size, dome-shaped membranes, alignment grooves and a metal mesh cover \cite{Prodanovic2019}. First, the window diameter can be increased to $\SI{45}{\um}$, which improves the active surface area to $52.6\%$. Second, the dome-shaped membranes have a focusing effect on SEs: they are directed to the center of the next dome. This ensures that SEs are contained within their 'channel' above the corresponding pixel pad. The dome shapes are created by etching and smoothing circular pillars into bumps on the silicon substrate before depositing the film layer. Third, alignment grooves are etched in the silicon frame of the tynode, which are used to align two tynodes by placing glass quartz wires between them. This 'self-alignment process' improves the accuracy of the alignment. Last, the dielectric mesh can be covered with a metal layer, which will prevent electrons to be trapped. Although, these new features alleviate some of the concerns, it increases the complexity of the fabrication process.
\\
\\
In this paper, we present an entirely different tynode design, which eliminates the need of a supporting mesh by forming a corrugated membrane of ultra-thin films. In a recent paper, a new class of nanoscale mechanical metamaterials with enhanced robustness, flexibility, rigidity and strength was reported by Davani et al. \cite{Davami2015}. They formed corrugated plates by atomic-layer-deposition (ALD) of an ultra-thin aluminum oxide film on a patterned silicon wafer, which acted as a mold and was removed afterwards. Their plate metamaterials were extremely flat, ultra-light and had shape-recovering properties. By using a corrugated membrane, the mechanical strength is enhanced compared to a flat membrane, while the membrane thickness is constant throughout the entire structure. The corrugated Al\textsubscript{2}O\textsubscript{3} film can be functionalized as a tynode by adding a conductive layer, such as titanium nitride \cite{Prodanovic2018,Chan2020}.

This design has many advantages. First, the effective area of these films is nearly 100\%: an incoming electron can generate transmission SEs on any part of this ultra-thin corrugated membrane. Second, alignment precision is less stringent in this case, so the inclusion of alignment grooves becomes optional. Third, the honeycomb-shaped domes can be tailored to have a focusing effect, which is still needed to direct SEs onto the pixel pads. Fourth, the risk of charge accumulation within the tynode stack is eliminated, since the thick dielectric mesh is no longer present. Last, the fabrication process is less complex in comparison. 

A metamaterial is defined as a material engineered to have properties that does not naturally occur. The corrugated membrane presented in this work is categorized as a mechanical metamaterial, but the multi-layered film is also engineered to sustain secondary electron emission. In this paper, we will refer to 'metamaterial' as a material with enhanced mechanical strength and improved secondary electron emission properties. The corrugated membranes of multi-layered Al\textsubscript{2}O\textsubscript{3}/TiN/Al\textsubscript{2}O\textsubscript{3} films are referred to as 'metamaterial tynodes'. In figure \ref{fig:2}, a schematic drawing of a TiPC detector with metamaterial tynodes is drawn, which has a continuous surface for electron multiplication. In this paper, metamaterials with two different patterns are presented: one with a hexagonal/honeycomb pattern and one with an octagonal pattern. The latter is designed to match the pitch of the pixel pads of a TimePix chip. The electron emission yield of these metamaterial tynodes will be determined within a scanning electron microscope using the collector-based method reported in Ref. \cite{Chan2020}. 

	
\section{Design}
\label{sec:design}

\begin{figure}
\centering 
\includegraphics[width=0.9\textwidth,origin=c,angle=0]{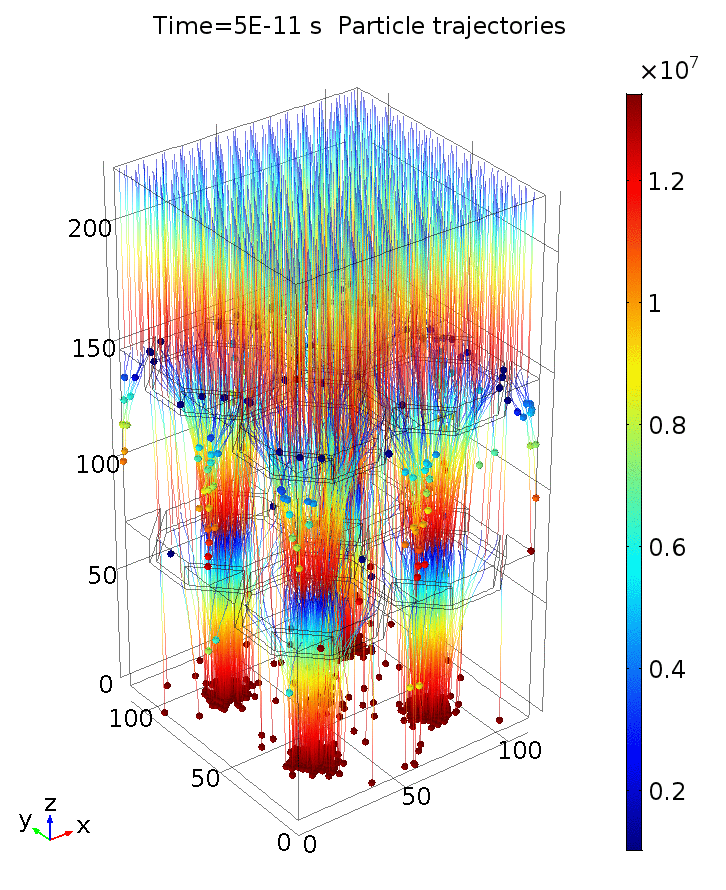}
\caption{\label{fig:3} The corrugated membranes are suspended within a silicon frame with a dimension of $\SI{20}{\mm}$ by $\SI{20}{\mm}$. For the hexagonal pattern, 16 square windows are opened in the silicon frame. Each corrugated membrane has an active area of $\sim\SI{1}{\square\mm}$. For the octagonal pattern, the silicon frame has a single window in which a membrane with an active area of $\sim\SI{16}{\square\mm}$ is suspended. Both patterns meet the design requirement that any perpendicular plane (green lines) is intersected by a vertical wall (black lines).}
\end{figure}

\begin{figure}
\centering 
\includegraphics[width=0.5\textwidth,origin=c,angle=0]{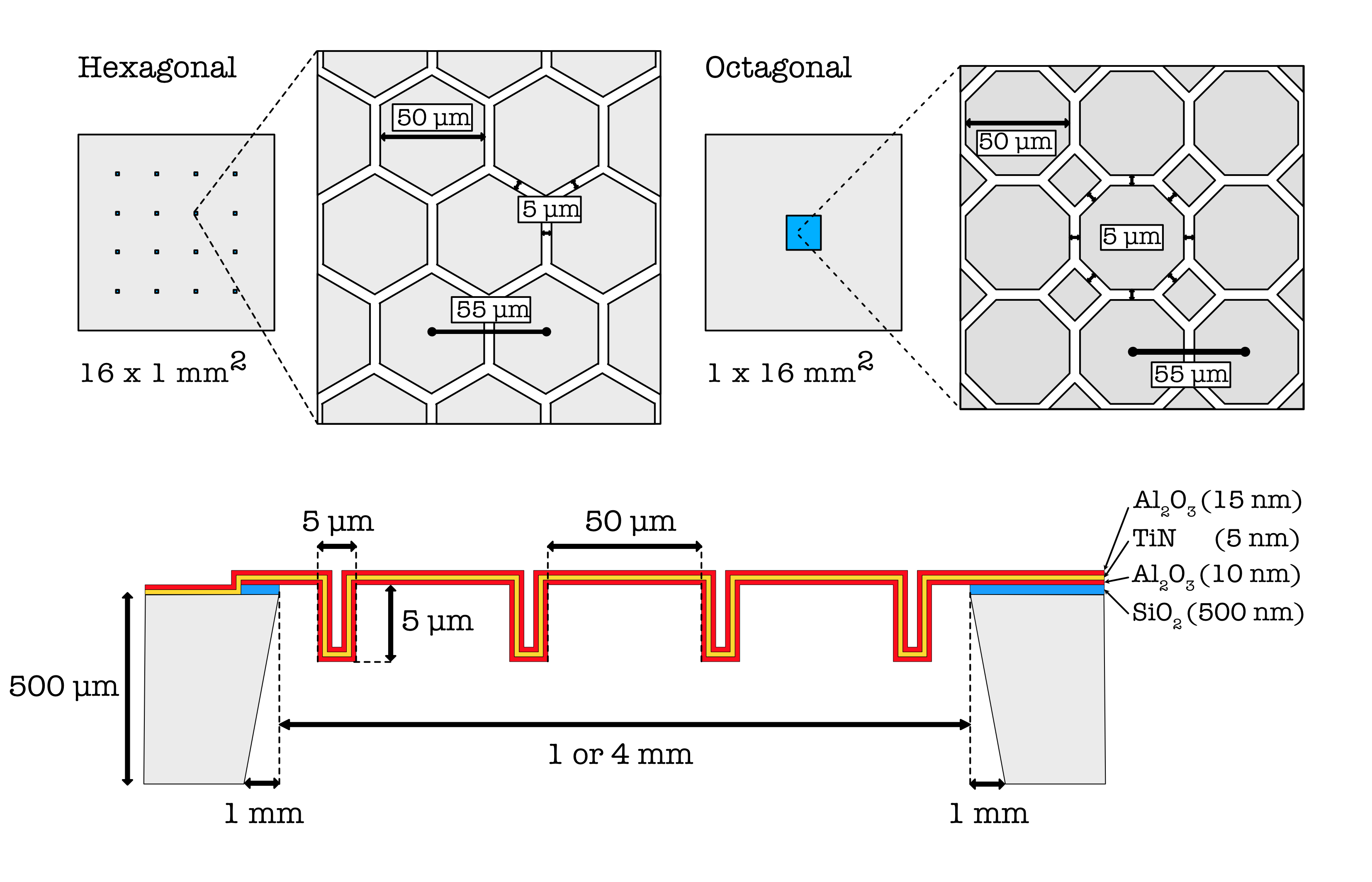}
\caption{\label{fig:4} COMSOL simulation of a TiPC detector with two tynode stages. For the simulation, two modules of COMSOL were used: AC/DC and Charged Particle Tracing (CPT). The first module is used to simulate the static electric field. The potential difference $\Delta U$ between each stage is $\SI{1000}{\V}$. The second module is used to simulate transmission secondary electron emission from the two tynode stages. Photoelectrons are released uniformly from the photocathode. Each incoming electron generates one transmission secondary electron on each tynode. Each PE and the SEs that they generate on each tynode will form a 'path', enabling us to estimate the collection efficiency of a TiPC detector.}
\end{figure}

The mechanical plate metamaterial reported by Davami et al. has unique mechanical properties: enhanced robustness, flexibility, bending stiffness and flatness \cite{Davami2015}. The enhanced robustness stems from the hexagonal honeycomb pattern. A failure mode of flat membranes due to stress is the formation of cracks, which propagate in straight lines. The triangular lattice of the honeycombs prevents this and localizes defects. Therefore, the chance of rupture is reduced. The enhanced bending stiffness and rigidity is provided by the intersections of the vertical walls and the horizontal segments, which resist deformation, although at the same time the flexibility is enhanced due to the non-continuous top and bottom surface, which allow the plates to be folded (figure \ref{fig:3}). The combination of enhanced bending stiffness and flexibility leads to the shape-recovery property of the plates after extreme deformation. 

Davami et al. also discussed design rules for their plate metamaterial \cite{Davami2015}. First, a periodic pattern must be chosen, which for any plane perpendicular to the plate intersects with a vertical wall (figure \ref{fig:3}).  Otherwise, the plate will bend along a line that only contains horizontal elements. Second, the enhancement factor of the bending stiffness depends on the diameter of the cell, rib width and rib height. Although, the bending stiffness saturates for heights above $\SI{1}{\um}$. The highest bending stiffness can be achieved by minimizing the rib width $w$ and maximizing the cell diameter $D$. In that case, the enhancement factor is given by $EF_\text{max} \approx (\frac{D}{w})^2$. Although, a large cell diameter will reduce the robustness of the plate due to the increased risk of crack propagation on the cell surface.

With these design rules in mind, two patterns are considered for the metamaterial tynodes: hexagonal honeycomb and octagonal. The hexagonal pattern fulfils the design criteria and has the desired properties, but the triangular lattice does not match the square spacing of pixels on a TimePix chip. Therefore, the octagonal pattern is designed to be compatible with a TimePix chip. First, the square lattice of the pattern matches the pitch/displacement of the pixel pads. Second, the octagonal-shaped cups have a focusing effect on SEs, which concentrate and bundle SEs above the individual pixel pads. The focusing effect of these octagonal-shaped cells has been simulated with COMSOL (figure \ref{fig:4}). A cell with a height of $\SI{5}{\um}$ is already sufficient to focus SEs, but the focusing point can be tailored by varying the height of the unit cells and/or the potential difference between the multiplication stages. Although, the latter is often attuned to the required electron energy to obtain the highest TEY for a given film thickness and is therefore a less flexible design parameter. The simulation also shows that PEs and SEs that enter the small square cups in the corner of each octagonal tend to land next to the pixel pads. This needs to be taken into consideration when estimating the collection efficiency. 

The die size of the samples is 2 by \SI{2}{\cm} in which windows are opened by deep-reactive ion etching (\gls{DRIE}) etching as shown in figure \ref{fig:3}. For the hexagonal honeycomb pattern, there are 16 square windows with a width of $\SI{1.25}{\mm}$. For the octagonal pattern, there is a single square window with a width of $\SI{4.2}{\mm}$. These window sizes are defined on the back side of the silicon wafer, but will be approximately $\SI{0.2}{\mm}$ smaller on the front side after etching since the DRIE process is not entirely anisotropic. The active surfaces are approximately $\SI{1}{\square\mm}$ and $\SI{16}{\square\mm}$, for the hexagonal and octagonal pattern respectively. Each unit cell in the corrugated membrane has a rib height and width of $\SI{5}{\um}$, which is relatively shallow and open. Incoming electrons can still reach the bottom of the rib and contribute to electron multiplication. The ratio of 1:1 is ideal, since increasing the height further would create deep trenches for which it is more difficult for electrons to reach the bottom, i.e. the angle of incidence becomes smaller. Also, the height is sufficient in this case for focussing. The cell diameter is $\SI{50}{\um}$ and has a pitch of $\SI{55}{\um}$, which is the same pitch between pixels on a TimePix chip. The film is a tri-layer Al\textsubscript{2}O\textsubscript{3}/TiN/Al\textsubscript{2}O\textsubscript{3} composite with a thickness of 10/5/$\SI{15}{\nm}$. This multilayered film is engineered as a membrane for electron multiplication by encapsulating a TiN layer, which provides in-plane conductivity that is needed to sustain electron emission \cite{Chan2020}.


\section{Fabrication}
\label{sec:fab}

\begin{figure}
  \centering
	\subfloat[Silicon mold. A 3D pattern is etched on the front side of a silicon wafer by DRIE. \label{fig:5a}]{\includegraphics[scale=0.35]{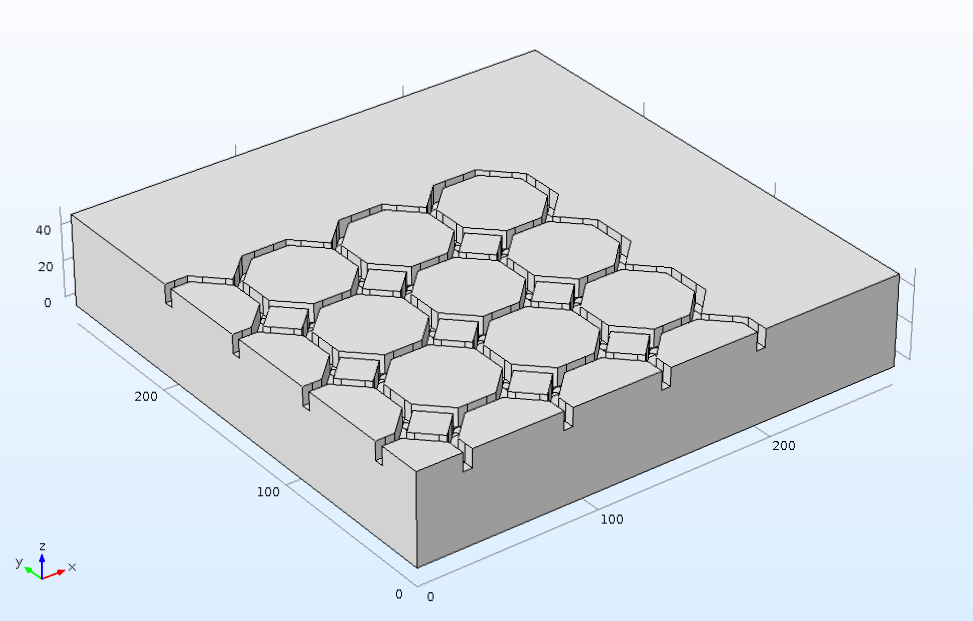}} 
	\qquad
	\subfloat[ALD of a tri-layer film. The Al\textsubscript{2}O\textsubscript{3}/TiN/Al\textsubscript{2}O\textsubscript{3} film is deposited uniformly on the 3D pattern.\label{fig:5b}]{\includegraphics[scale=0.35]{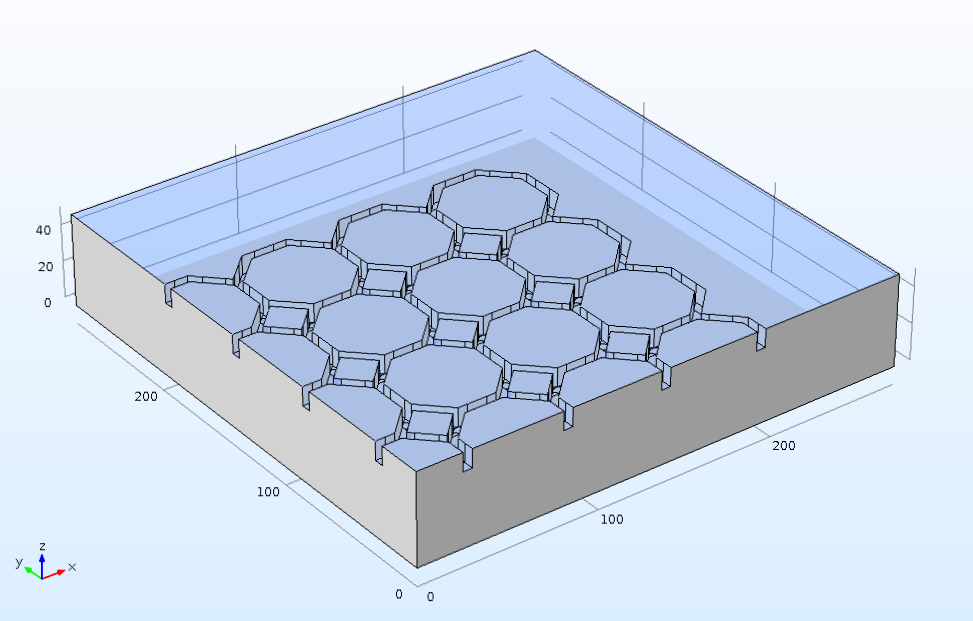}} \\
	\subfloat[Release of the metamaterial film. On the backside of the wafer, square windows are etched into the silicon substrate by DRIE. The landing layer is a silicon dioxide film, which is removed by HF vapor etching afterwards. \label{fig:5c}]{\includegraphics[scale=0.35]{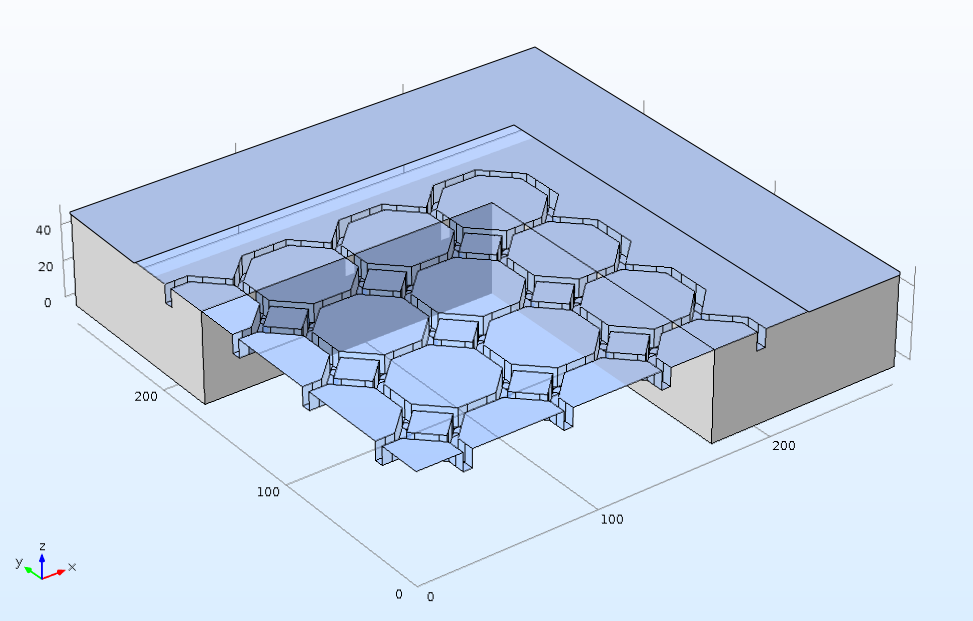}} \
     \caption{Overview of the fabrication process} \label{fig:5}
\end{figure}	

Wafers (4-inch, p-type with 5-$\SI{10}{\ohm\cm}$) with a thickness of $500\pm \SI{15}{\um}$ are used as a substrate. The wafers are cleaned with a standard cleaning procedure before alignment markers are etched on the wafers for contact alignment. The sequence of the cleaning procedure is as follows: a plasma oxygen etch, a HNO\textsubscript{3} 100\% bath for $\SI{10}{\min}$, a demineralized water rinse for $\SI{5}{\min}$, a HNO\textsubscript{3} 65\% bath at $\SI{110}{\celsius}$ for $\SI{10}{\min}$ and another water rinse. 

The process can be divided in three parts. In the first part, a 3D mold is etched within the silicon substrate (figure \ref{fig:5a}). A 3-\SI{}{\um}-thick photoresist (\gls{PR}) layer is used as a masking layer for a deep-reactive ion etch in a Rapier Omega i2L DRIE etcher. The hexagon/octagon pattern is transferred to the PR and trenches with a depth of $\SI{5}{\um}$ are etched into the silicon substrate. The wafers are then cleaned with oxygen plasma to remove residual polymers from the DRIE process followed by a standard cleaning procedure. Afterwards, the wafers are put in an oven at $\SI{1100}{\celsius}$ to wet thermally grow a silicon dioxide layer with a thickness of $\SI{500}{\nm}$. This layer will act as a sacrificial and stopping layer. 

In the second part, the tri-layer film material is deposited conformally on the mold (figure \ref{fig:5b}). First, a layer of Al\textsubscript{2}O\textsubscript{3} is deposited by atomic-layer-deposition (ALD) in a thermal ALD ASM F-120 reactor at $\SI{300}{\celsius}$ using trimethyl-aluminum (TMA) and water as a precursor and reactant, respectively. A strip of the newly deposited layer is removed from the edge of each die to create a contact point for the next layer to the silicon substrate. First, the Al\textsubscript{2}O\textsubscript{3} layer is plasma etched in an Omega Trikon plasma etcher. Then, the silicon dioxide layer is removed with a plasma etch in a Drytek plasma etcher. The wafers are cleaned afterwards using a new cleaning procedure that omits the 'fuming' HNO\textsubscript{3} 65\% bath step. Next, a layer of ALD titanium nitride is deposited in an Ultratech Fiji G2 using titanium chloride (TiCl\textsubscript{4}) as precursor and nitrogen plasma as reactant at $\SI{250}{\celsius}$. The third layer of Al\textsubscript{2}O\textsubscript{3} is then deposited with the same process parameters as the first layer in the same reactor. 

In the third part, the corrugated membrane will be released by opening windows in the silicon substrate (figure \ref{fig:5c}). First, a plasma-enhanced chemical vapor deposition (\gls{PECVD}) oxide layer with a thickness of $\SI{1}{\um}$ is deposited on the front side of the wafer in a Novellus Concept One system. This oxide layer protects the tri-layer film from mechanical damages, since the wafers are placed upside down in the subsequent steps. The backside of the wafer is stripped by using two plasma etches: the ALD TiN and ALD Al\textsubscript{2}O\textsubscript{3} layers are stripped in an Omega Trikon plasma etcher, while the thermal oxide is removed in a Drytek plasma etcher. Once the backside is stripped and cleaned, a PECVD oxide layer with a thickness of $\SI{5}{\um}$ is deposited on the backside of the wafer in a Novellus Concept One system. This PECVD oxide layer is the masking layer for DRIE. The pattern of the window openings and scribe lines are transferred to the backside of the wafer using PR with a thickness of $\SI{3.1}{\um}$. This pattern is then plasma etched into the PECVD oxide masking layer in a Drytek plasma etcher. 
	The wafers are now ready for the final release using a Rapier Omega i2L DRIE etcher. First, a fast DRIE recipe is used to etch approximately $\SI{495}{\um}$ into the silicon substrate until the rib pattern of the corrugated film becomes visible. This is one of the critical steps in this process, since over-etching can damage the structure of the ribs of the corrugated membrane. The protective PECVD oxide on the front side is then removed in a hydrogen fluoride (HF) vapor etch chamber using 4 etching cycles of HF and ethanol with a flow of 190 sccm and 220 sccm, respectively, for a duration of 10 min per cycle. If this oxide layer is not removed before the final release steps, it will be the main contributor to the stress within the film and can cause ruptures. The wafer is then cleaved along the scribe lines. The individual dies are transferred to a carrier wafer with pockets to hold them. The remaining $\SI{5}{\um}$ of silicon is removed with a slower DRIE recipe within the Rapier. At this point, the dies become fragile and should be handled with care. The dies are then cleaned with oxygen plasma in a TEPLA plasma cleaner to remove any residual polymers from the DRIE process. The final and critical step is the removal of the thermal oxide layer in a HF vapor etch chamber using the same recipe as before. This step is repeated until the ultra-thin membranes are released. For the samples presented in this work, the recipe was repeated 3 times before the membranes were successfully released. The optimal etch time for these dies was ${4\times}\SI{10}{\min}$. After the final release, the membranes are fragile (but still relatively strong considering their thinness) and should be transported with care.


\section{Experimental setup}
\label{sec:setup}

\begin{figure}
  \centering
	\subfloat[\label{fig:6a}]{\includegraphics[height=7.5cm]{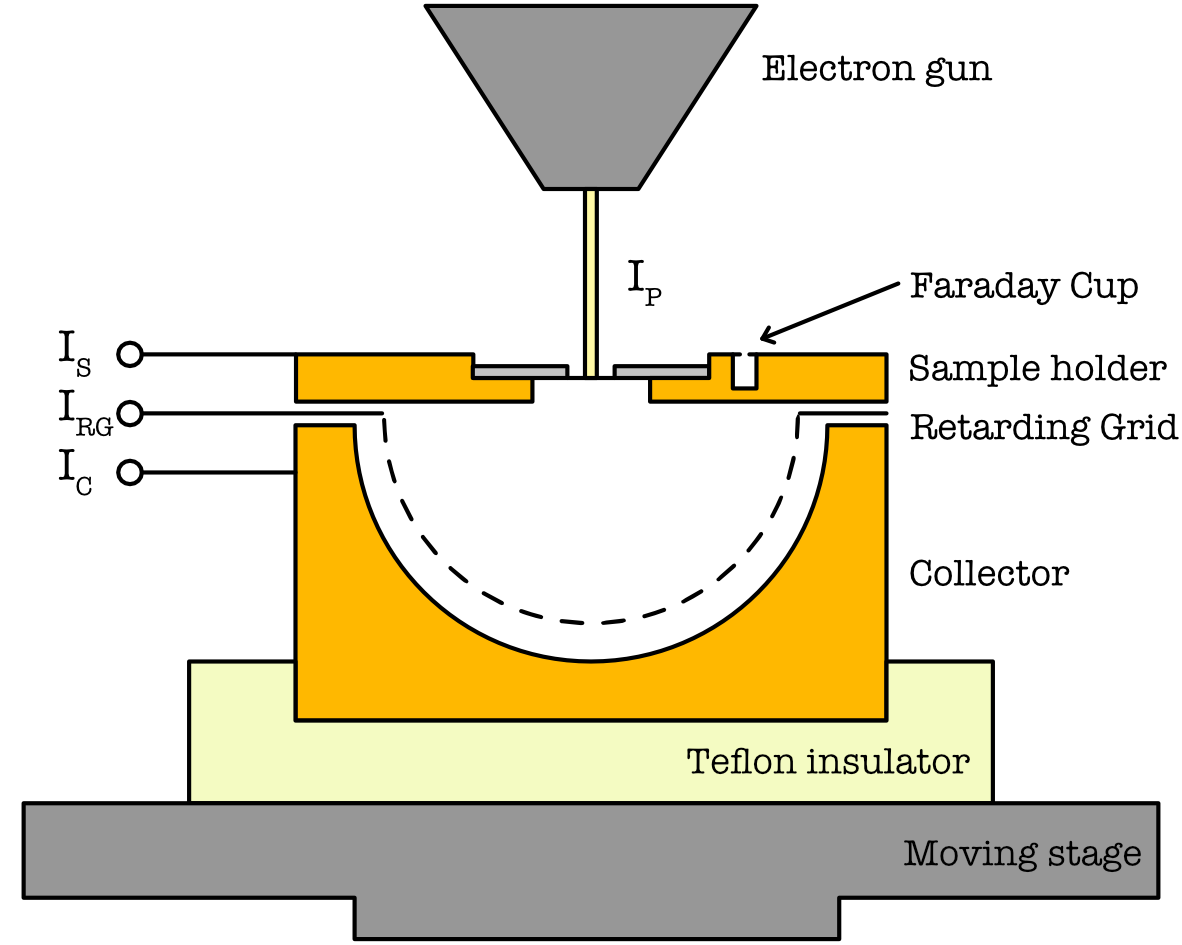}} \
	\subfloat[\label{fig:6b}]{\includegraphics[height=7.5cm]{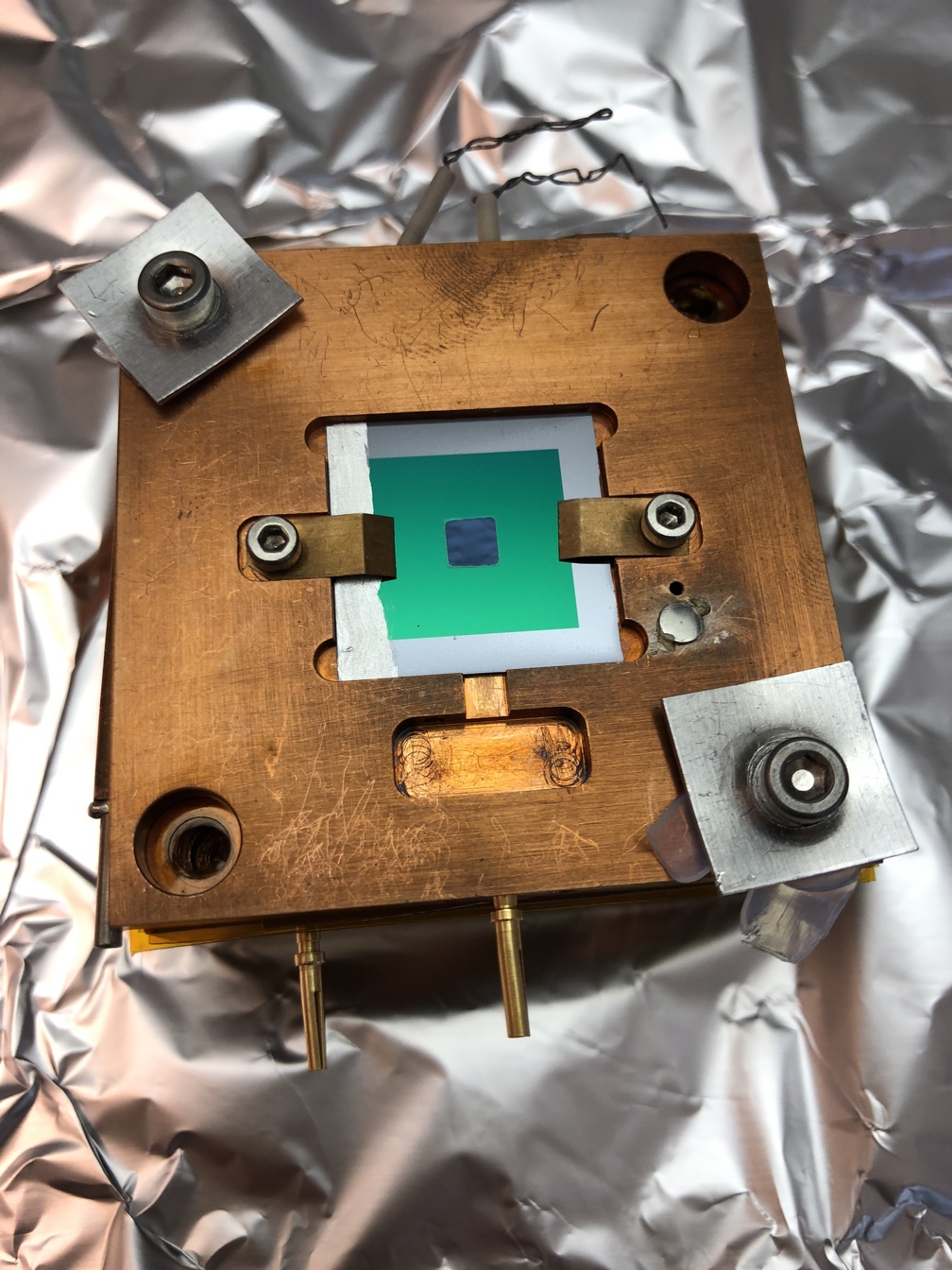}}
     \caption{Experimental setup. \protect\subref{fig:6a} schematic drawing of the collector system. The forward-scattered electrons (\gls{FSE}s) and transmission secondary electrons (\gls{TSE}s) are collected by the retarding grid and/or collector. By measuring the retarding grid current $I_{RG}$ and the collector current $I_C$, the FSE and TSE yields can be determined. The backscattered electron (\gls{BSE}) and reflection secondary electrons (\gls{RSE}) yields are determined indirectly by measuring the sample current $I_S$. \protect\subref{fig:6b} The collector system with one of the octagonal metamaterial membrane mounted in the sample holder.} \label{fig:6}
\end{figure}	

\subsection{Transmission secondary electron yield}
\label{ssec:yield}

The transmission (secondary) electron yield is measured with a hemispherical collector system in a scanning electron microscope (\gls{SEM}) as shown in figure \ref{fig:6a}. A more extensive description of the method is given in ref \cite{Chan2020}. The system is mounted on the moving stage of a Thermo Fisher NovaNanoLab 650 Dual Beam SEM, which allows us to use all functions of the SEM while performing TEY measurements. The sample holder, mesh grid and collector are connected to Keithley 2450 source meters via a feedthrough into the vacuum chamber. Before the measurements, the electron beam current is measured using a Faraday cup that is placed next to the sample. The PE energy ranged from 0.3 – $\SI{10}{\keV}$ with a beam current from 0.06 to $\SI{0.54}{\nA}$. The beam current $I_0$ depends on the PE energy and is measured for each energy. 

The measurement is performed with a scanning electron beam by using the image acquisition mode of the SEM. The raster pattern distributes the electrons over a larger surface and lowers the electron dose. This mitigates charge-up effects and/or the build-up of surface contamination. The following SEM settings are used during the measurement: dwell time of $\SI{1}{\us}$, magnification of 50X, horizontal field width of $\SI{2.56}{\mm}$ and a resolution of 1024 x 884. The dwell time is defined as the average time that the electron beam is irradiating the specimen to acquire one pixel in a SEM image. Using these settings, the area of the irradiated surface can be estimated, which in this case is $\SI{5.66}{\square\mm}$. The measured yields are averaged over the surface and over time, i.e. the yield is calculated from multiple frames. For the measurement of the smaller membranes with the hexagonal pattern, a horizontal field width of $\SI{0.640}{\mm}$ is used.

Primary electrons can generate backscattered electrons (\gls{BSE}s), reflection secondary electrons (\gls{RSE}s), forward-scattered electrons (\gls{FSE}s) and transmission secondary electrons (\gls{TSE}s) from a tynode (figure \ref{fig:6a}). The transmission currents are measured directly within the collector, while the reflection currents are determined indirectly using the sample current. The fast electrons (BSEs and FSEs) can be separated from the slow electrons (RSEs and TSEs) by performing a measurement with a negative sample bias ($-\SI{50}{\V}$) and a measurement with a positive sample bias ($+\SI{50}{\V}$). The Keithley 2450 sourcemeters can perform current measurements while simultaneously applying different biases to the electrodes. When a negative bias is applied to the sample, the sample holder, retarding grid and collector are biased at -50, 0, -50 V, respectively. When a positive bias is applied to the sample, they are at +50, 0, +50V. In the first case, both fast and slow electrons are repelled from the tynode. The total transmission yield $\sigma_T(E_0)$ is given by 
\begin{equation}
\label{eq:1}
\sigma_T(E_0) = \frac{I_{RG-}+I_{C-}}{I_0}
\end{equation}
where $E_0$ is the electron energy of the electron beam, $I_0$ is the beam current, $I_{RG-}$ is the retarding grid current and $I_{C-}$ is the collector current. The subscript '-' indicates that the currents are measured with a negatively biased sample. The total emission yield is given by: $\sigma(E_0)=(I_0-I_{S-})/I_0$. It is the sum of the total reflection and total transmission yield: $\sigma(E_0)=\sigma_T(E_0)+\sigma_R(E_0)$. Therefore, the total reflection yield $\sigma_R(E_0)$ is given
\begin{equation}
\label{eq:2}
\sigma_R(E_0) = \frac{I_0-I_{S-}-I_{RG-}-I_{C-}}{I_0}
\end{equation}
where $I_{S-}$ is the sample current. In the second case, the sample is positively biased and only fast electrons are measured. The FSE yield $\eta_T(E_0)$ and the BSE yield $\eta_R(E_0)$ are respectively given by 
\begin{equation}
\label{eq:3}
\eta_T(E_0) =  \frac{I_{RG+}+I_{C+}}{I_0}
\end{equation}
and
\begin{equation}
\label{eq:4}
\eta_R(E_0) =  \frac{I_0-I_{S+}-I_{RG+}-I_{C+}}{I_0}
\end{equation}
where $I_{S+}$ is the sample current, $I_{RG+}$ is the retarding grid current and $I_{C+}$ is the collector current. The subscript '+' indicates that the currents are measured with a positively biased sample. By subtracting the contribution of the fast electrons from the total emission yield, the true secondary electron yields can be determined
\begin{equation}
\label{eq:5}
\delta_T(E_0) =  \frac{I_{RG-}-I_{C-}}{I_0}-\frac{I_{RG+}+I_{C+}}{I_0}
\end{equation}
and
\begin{equation}
\label{eq:6}
\delta_R(E_0) =  \frac{I_0-I_{S-}-I_{RG-}-I_{C-}}{I_0} - \frac{I_0-I_{S+}-I_{RG+}-I_{C+}}{I_0}
\end{equation}

\subsection{Surface scan \& Yield maps}
\label{ssec:map}

A transmission secondary electron yield map can be constructed by measuring the transmission current as a function of time. The TSE yield is calculated by using equation \ref{eq:5} and is mapped to the coordinate of the electron beam during image acquisition. When a (single) SEM image is taken, the (transmission) emission current will vary across the surface of the sample. A yield map will show the difference in electron emission and can be compared to the SEM image. The method is described in appendix \ref{sec:yieldmap} and operates on the same principles as SEM image construction. The SEM settings are chosen such that the Keithley 2450 sourcemeters are able to map the measured current (as a function of time) to the pixels in the SEM image. This method requires a much slower scan speed than the method presented in section \ref{ssec:yield}. Also, only one image frame is acquired. A dwell time of $\SI{1}{\ms}$ is used, which is a compromise between speed and accuracy. A larger dwell time might cause charge-up effects and/or surface contamination. The resolution of the image is 512 x 442 acquired with a dwell-, line- and frame time of $\SI{1}{\ms}$, $\SI{560}{\ms}$ and $\SI{4.2}{\min}$ respectively. The electron beam energy used to acquire the image is $\SI{3.2}{\keV}$ with a current of $\SI{0.29}{\nA}$, which yields the highest transmission yield for the membranes considered in this paper. The magnification is 500X or 1000X, which shows the difference in yield across the 3D structure of the corrugated membrane more clearly.

\section{Results \& discussion}
\label{sec:results}

\subsection{Fabrication}
\label{ssec:resfab}

\begin{figure}
  \centering
	\subfloat[Front. Unit cells facing downwards $\downarrow$ \label{fig:7a}]{\includegraphics[width=7.5cm]{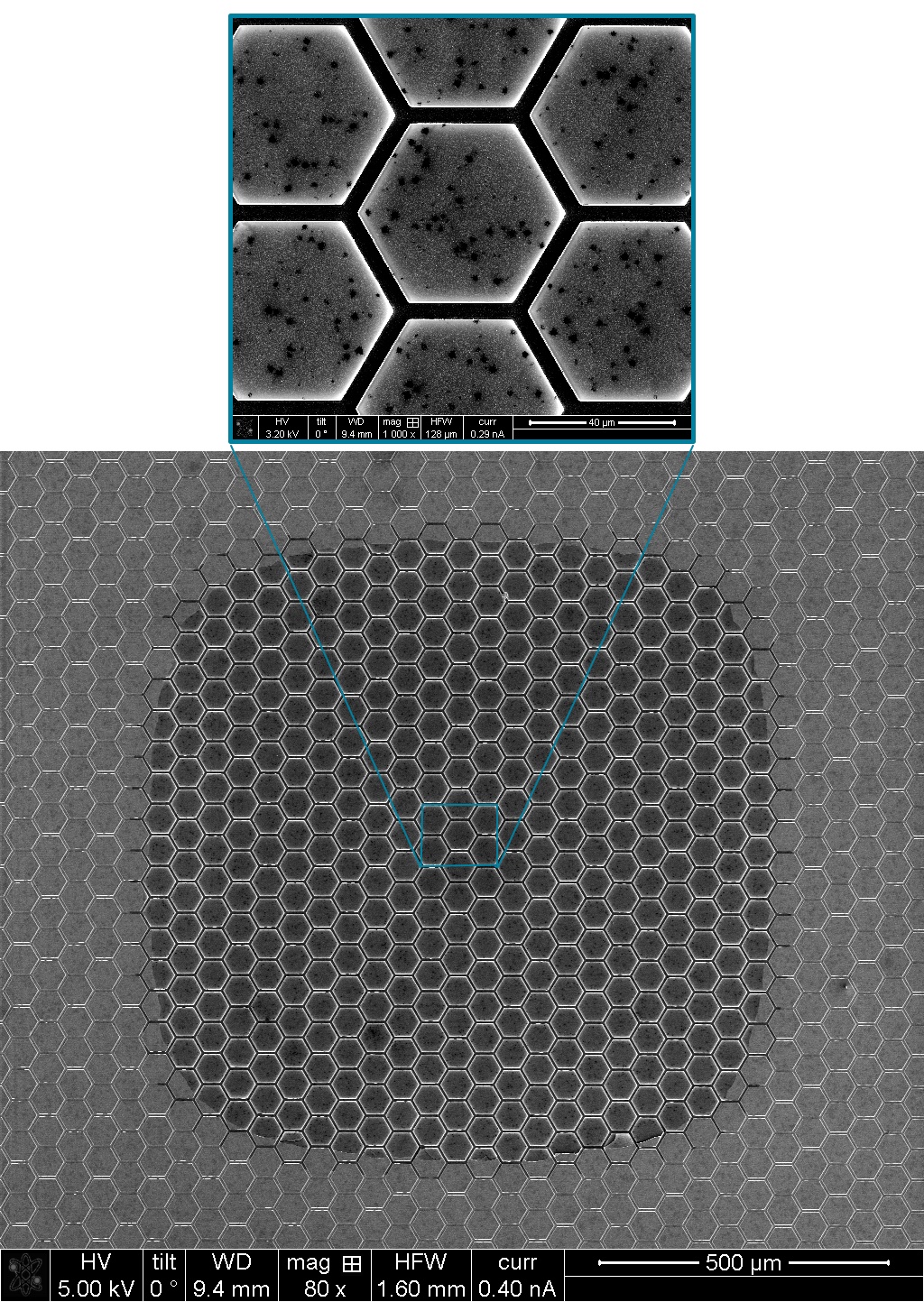}} \
	\subfloat[Back. Unit cells facing upwards $\uparrow$ \label{fig:7b}]{\includegraphics[width=7.5cm]{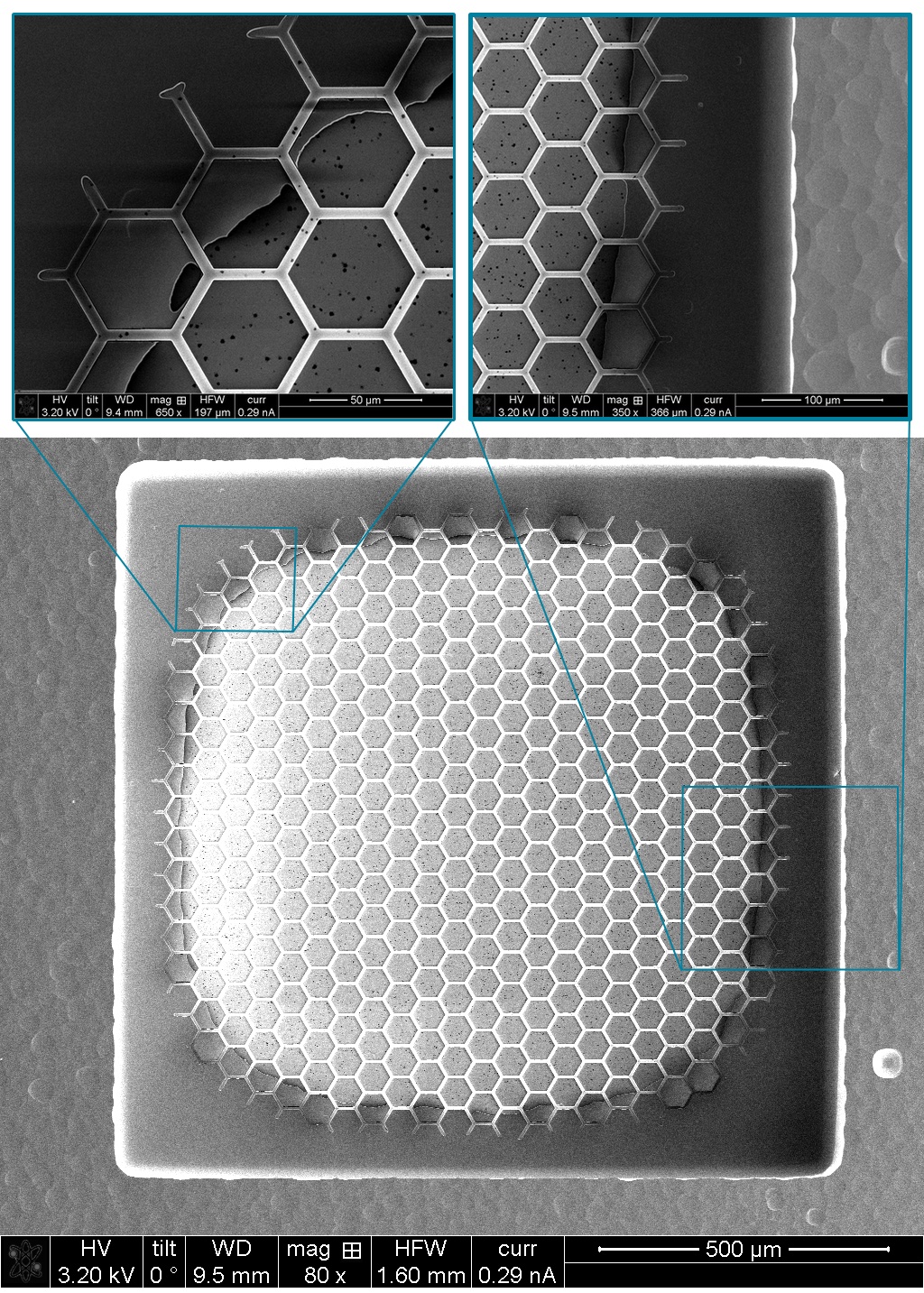}}
     \caption{SEM images of a metamaterial film with a hexagonal pattern. \protect\subref{fig:7a} The contrast between the active area and the window frame is due to the transparency of the film for 5 keV electrons. \protect\subref{fig:7b} On the backside, the ribs protrude outwards and appear brighter since also SEs emitted from the top and the sidewalls of the ribs can more easily reach the detector of the SEM. On the edge, the ribs disappear into the silicon substrate.} \label{fig:7}

\centering 
\includegraphics[width=7.5cm]{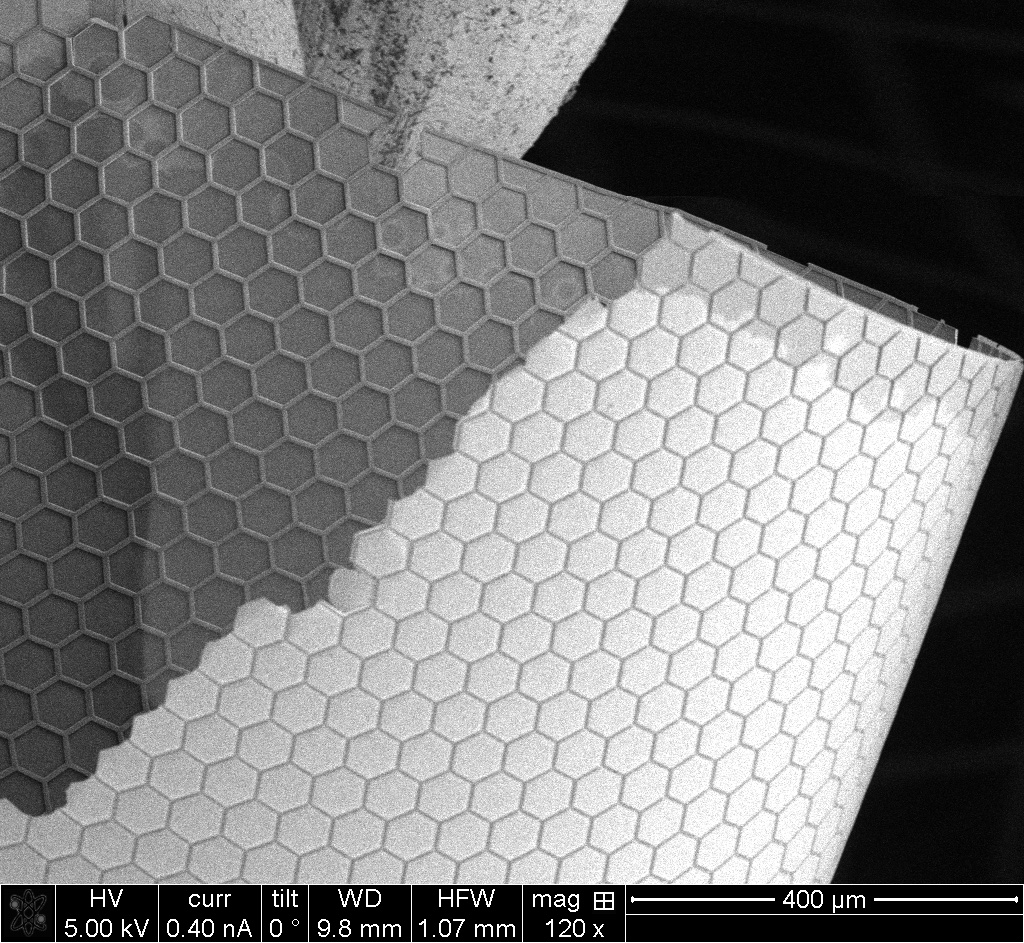}
\caption{\label{fig:8} A SEM image of a broken film curled up after release. It clearly shows the ribs of the honeycombs on the front- and backside.}
\end{figure}

\begin{figure}
  \centering
	\subfloat[Front. Unit cells facing downwards $\downarrow$ \label{fig:9a}]{\includegraphics[width=7.5cm]{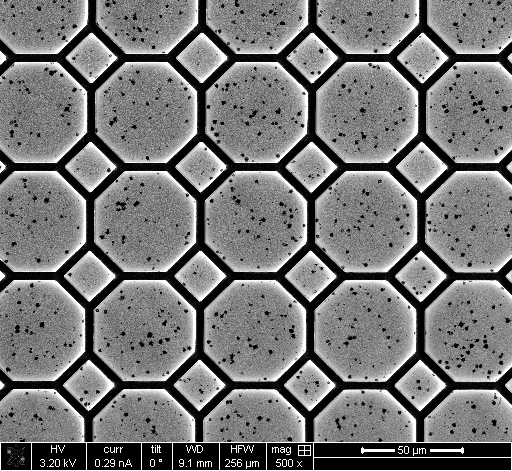}} \
	\subfloat[Back. Unit cells facing upwards $\uparrow$ \label{fig:9b}]{\includegraphics[width=7.5cm]{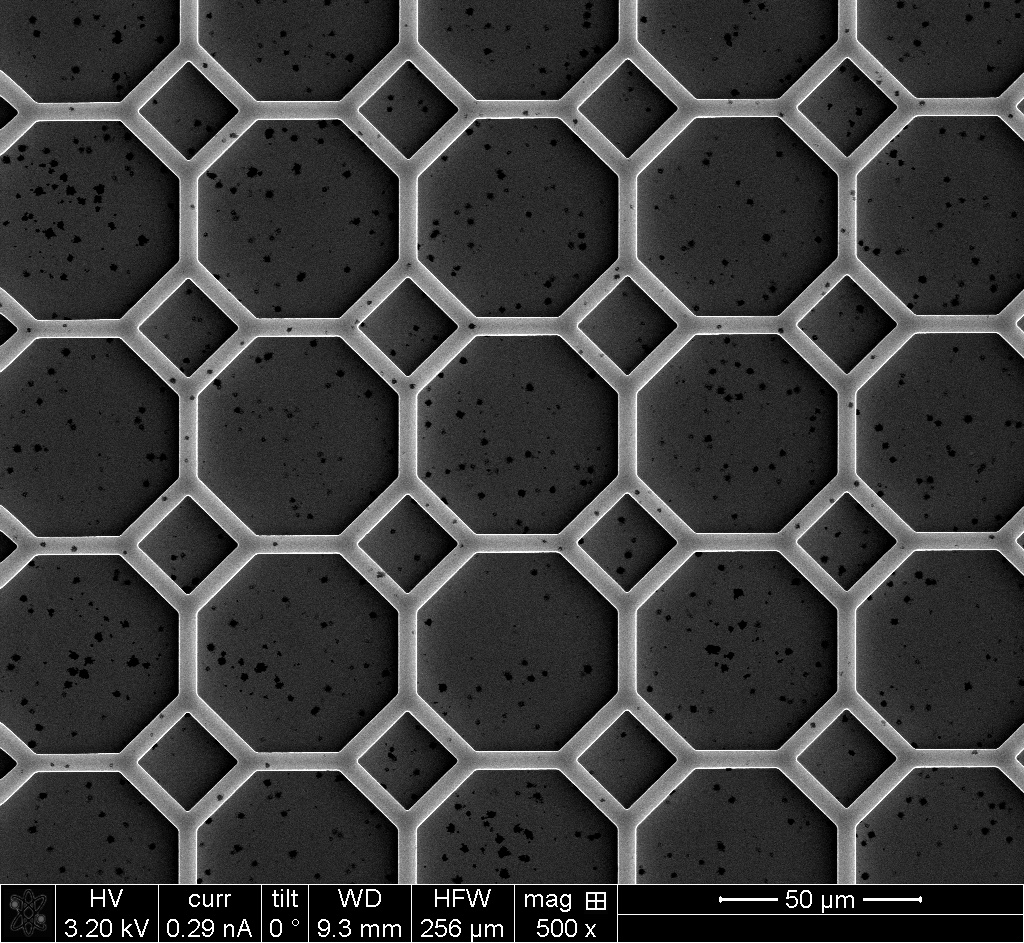}}
     \caption{SEM images of a metamaterial film with an octagonal pattern.} \label{fig:9}

  \centering
	\subfloat[\label{fig:10a}]{\includegraphics[width=7.5cm]{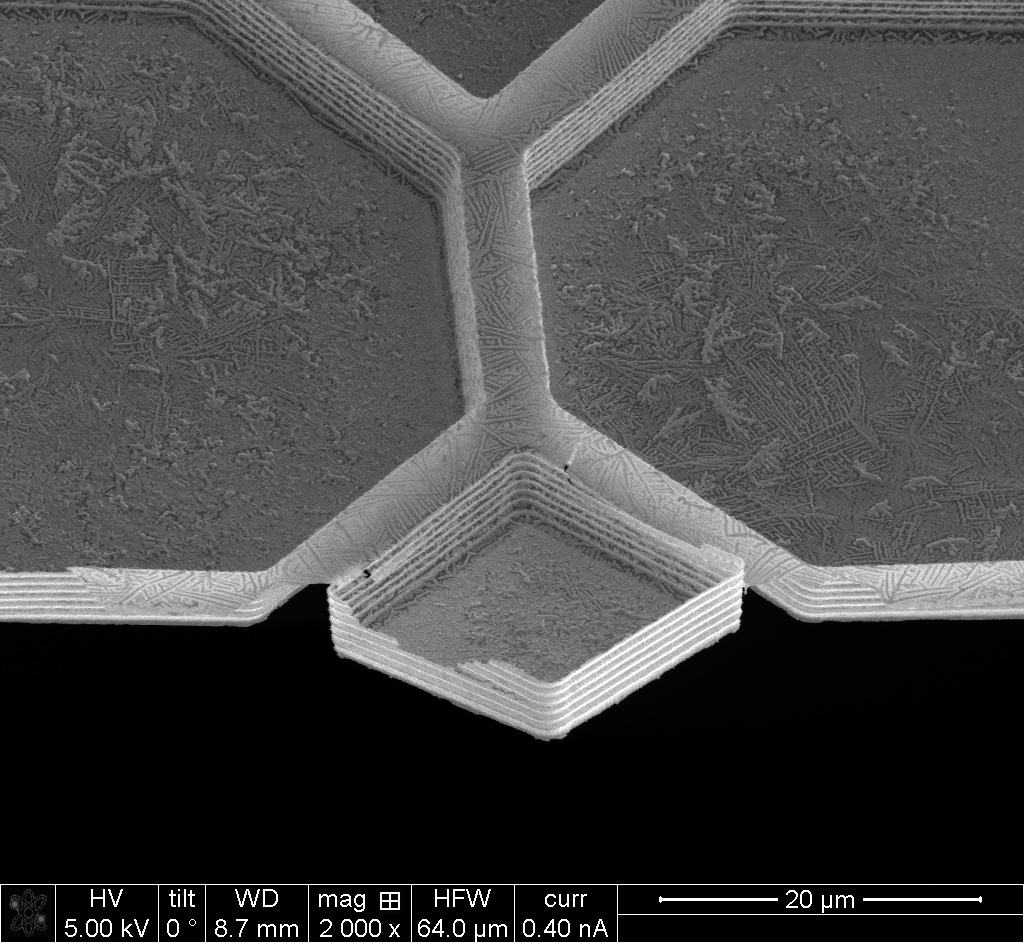}} \
	\subfloat[\label{fig:10b}]{\includegraphics[width=7.5cm]{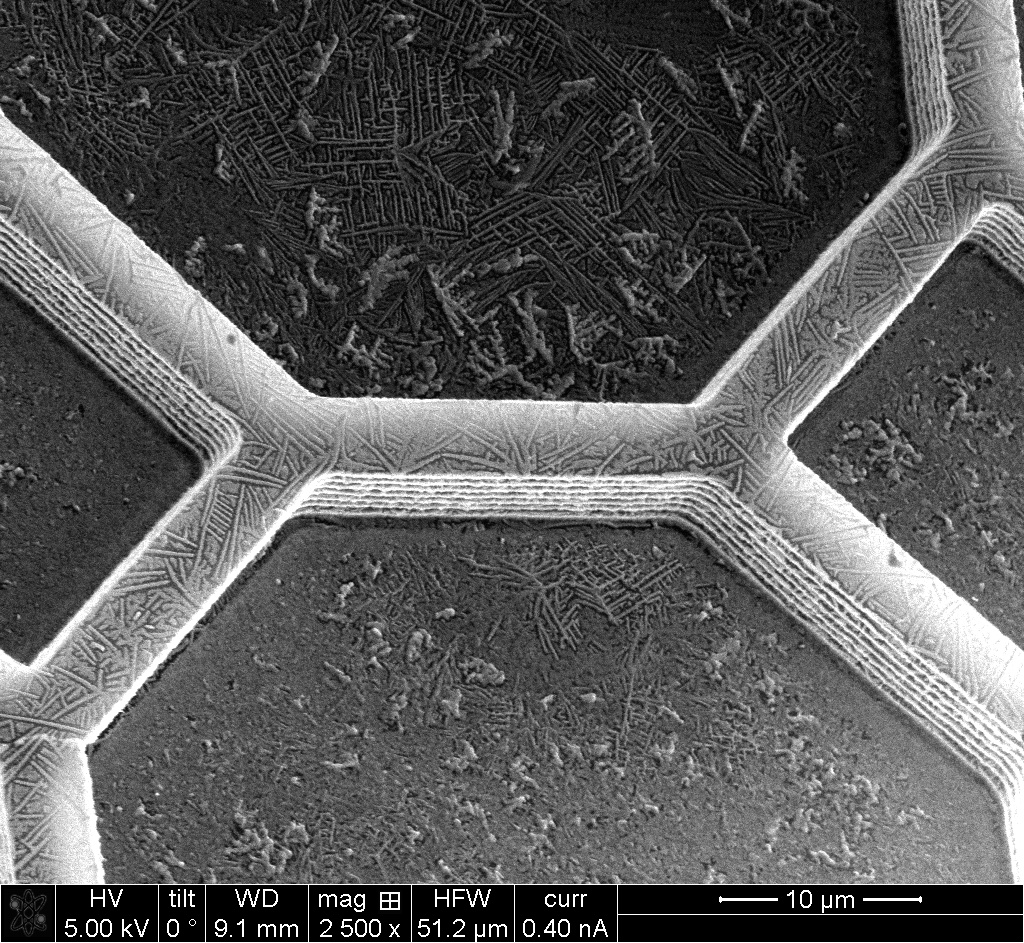}}
     \caption{Close-up of a broken metamaterial film with an octagonal pattern with the unit cells facing upwards $\uparrow$. The etch lines of the DRIE process are visible on the ribs.} \label{fig:10}
\end{figure}

In figure \ref{fig:7}, SEM images of the front- and backside of a metamaterial tynode with a hexagonal pattern are shown. The contrast in the image is due to the 3D structure of the film. In figure \ref{fig:7a}, the ribs extrude into the plane and behave as trenches from which it is difficult for SEs to escape, therefore the ribs are dark in the image. In figure \ref{fig:7b}, the backside of the film is shown. In this case, the ribs protrude out of the plane and are brighter than the film. The difference between both sides is more apparent in figure \ref{fig:8}, which shows a broken film that curled up after release. 

In figure \ref{fig:9}, SEM images of a metamaterial tynode with an octagonal pattern are shown. The black dots on the surface are residues from the fabrication process. On a different sample, a close-up of a broken film shows flakey residues on the film and the ribs (figure \ref{fig:10}). These are residues of the HF vapor etch. In figure \ref{fig:10b}, the etch lines of the DRIE process are visible as indentations in the ribs, which are due to the cyclic process of etching and passivation. For the samples in this paper, 8 cycles were used to etch $~\SI{5.1}{\um}$ into the silicon wafer. 

There are two critical steps in the fabrication process of these metamaterial tynodes. First, the DRIE process used to open the windows from the backside of the wafer requires careful monitoring. Although the thermal oxide layer acts as a landing layer, the etch rate is different for windows with different widths. In general, more open features are etched faster. As such, there is a chance of over etching, which can damage the ribs of the metamaterial film. Second, the metamaterial membranes are fragile after the final release step in the HF vapor chamber. They are prone to breakage due to air pressure differences and/or electrostatic forces. 

The largest window has a surface of $\sim \SI{16}{\square\mm}$ corresponding to 72 by 72 pixels, while a TimePix chip has 256 by 256 pixels. To span the entire chip, the window size should increase. However, the larger membranes will be more fragile. This can be remedied by making smaller windows and including a silicon frame. For instance, by sacrificing a row/column of 4 pixels (or a width of $\SI{0.22}{\mm}$), the surface of a TimePix chip can be covered by having 8x8 windows each accommodating 30x30 pixels. However, this solution decreases the collection efficiency.

\subsection{Transmission secondary electron yield}
\label{ssec:resTSEY}

\begin{figure}
  \centering
	\subfloat[Front. Unit cells facing downwards $\downarrow$ \label{fig:11a}]{\includegraphics[width=7.5cm]{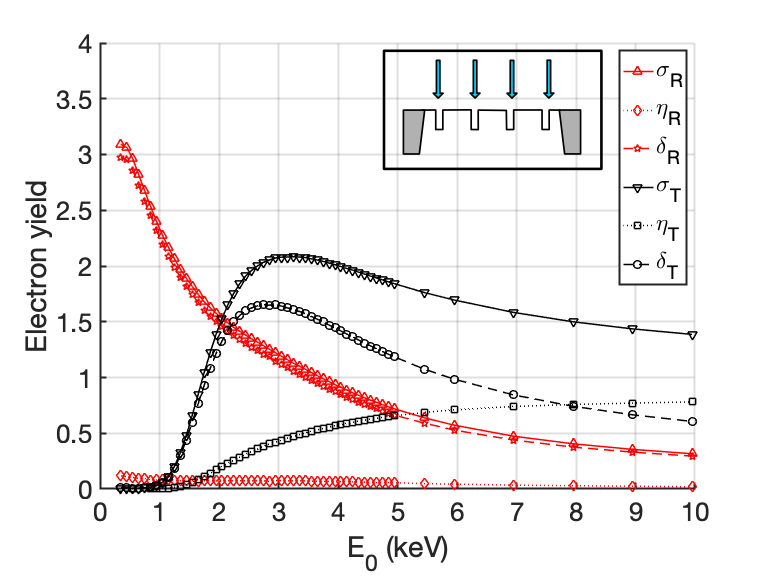}} \
	\subfloat[Back. Unit cells facing upwards $\uparrow$ \label{fig:11b}]{\includegraphics[width=7.5cm]{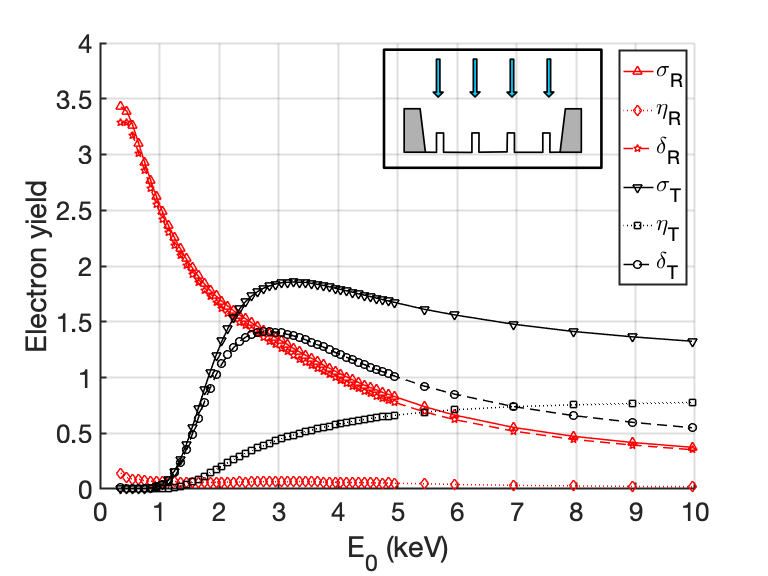}}
     \caption{Electron yield curves of an octagonal metamaterial film with a thickness of $15/5/\SI{10}{\nm}$. For each PE energy $E_0$, a surface with an area of \SI{5.75}{\mm^2} is irradiated by the electron beam and the average yield is calculated. Therefore, the variation in yield due to the microstructure of the corrugated film is averaged out. The resulting smooth yield curves are similar to the yield curves measured on flat membranes \cite{Chan2020}.} \label{fig:11}

  \centering
	\subfloat[Front. Unit cells facing downwards $\downarrow$ \label{fig:12a}]{\includegraphics[width=7.5cm]{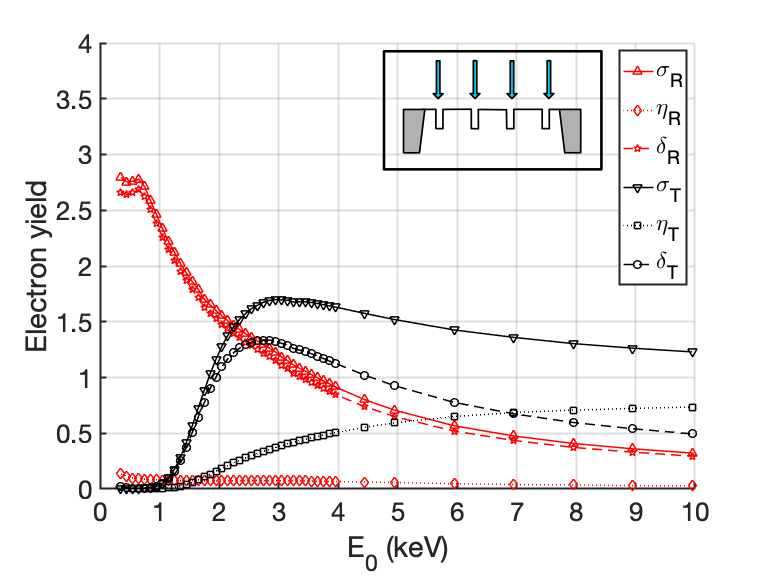}} \
	\subfloat[Back. Unit cells facing upwards $\uparrow$ \label{fig:12b}]{\includegraphics[width=7.5cm]{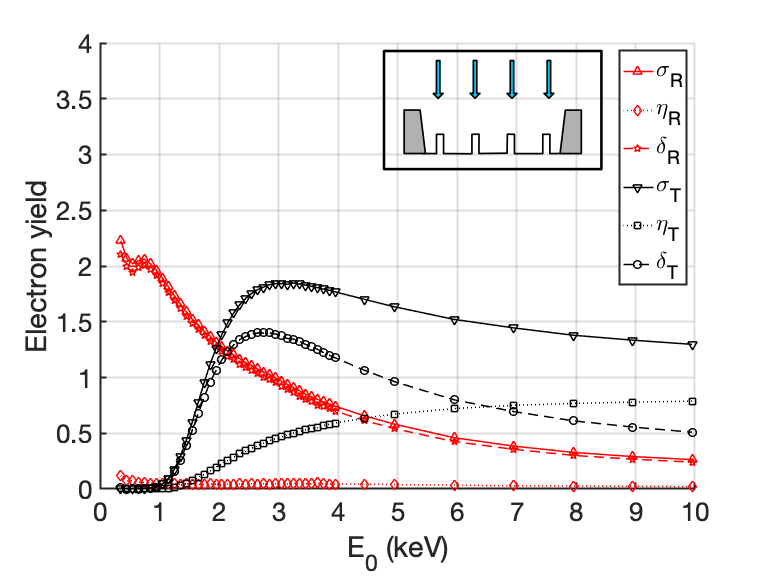}}
     \caption{Electron yield curves of an hexagonal metamaterial film with a thickness of $15/5/\SI{10}{\nm}$. For each PE energy $E_0$, a surface with an area of \SI{0.35}{\mm^2} is irradiated by the electron beam and the average yield is calculated. The window in which the hexagonal metamaterial film is suspended is smaller (see figure \ref{fig:3}).} \label{fig:12}
\end{figure}

\begin{figure}
\centering 
\includegraphics[width=9cm]{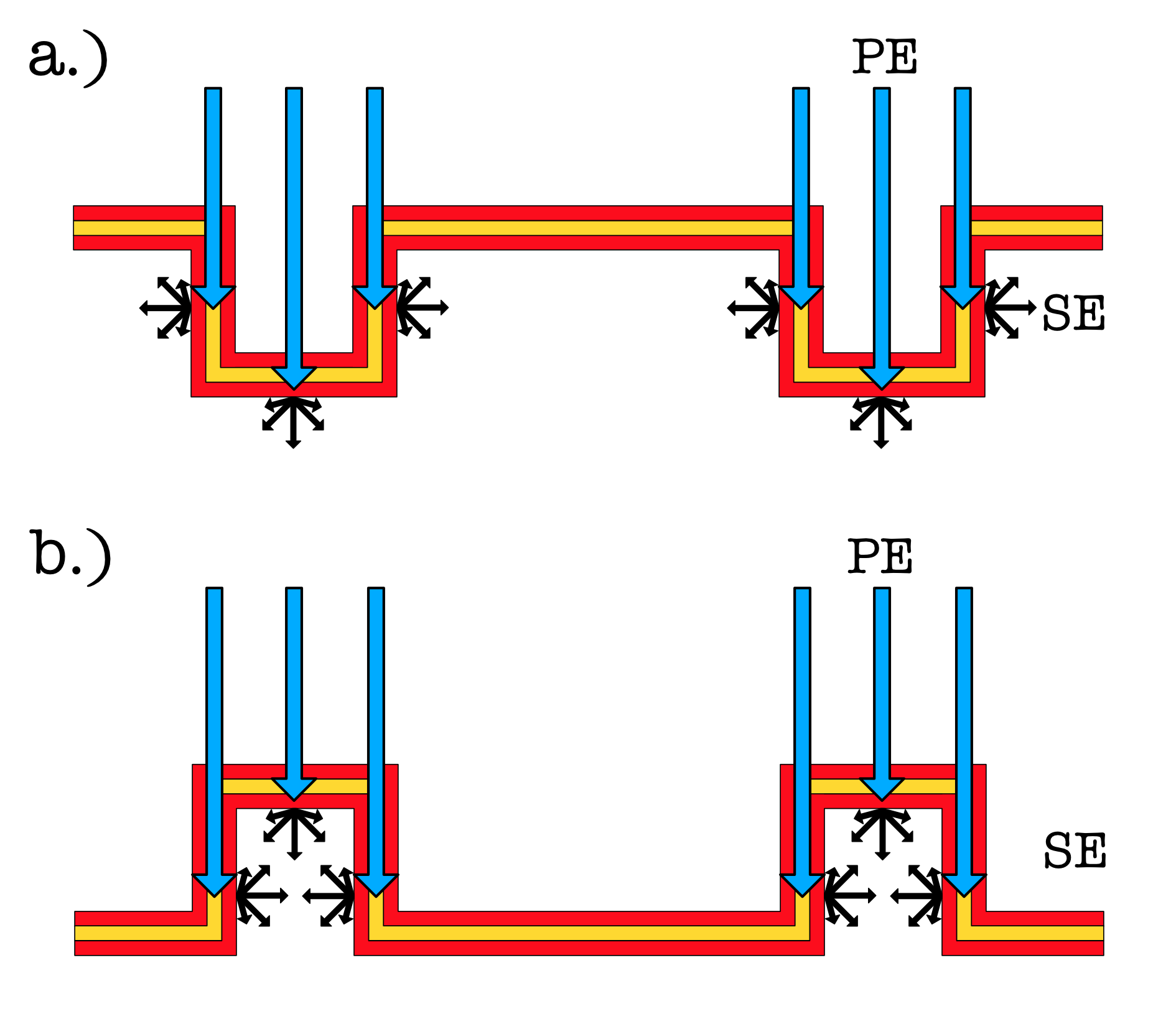}
\caption{\label{fig:13} The hexagon/octagon unit cell cups facing: a. downwards. TSEs escape into the unit cell cups and are focused to the next tynode. b. upwards. TSEs are recaptured within the rib.}
\end{figure}
	
In figure \ref{fig:11} and \ref{fig:12}, full sets of yields are given for the membranes with the octagonal and hexagonal pattern, respectively. A full set of yields consists of the BSE yield ($\eta_R$), the RSE yield ($\delta_R$), the FSE yield ($\eta_T$) and the TSE yield ($\delta_T$). The (total) transmission electron yield ($\sigma_T$) is the sum of the forward-scattered electron yield and the transmission secondary electron yield. It is the metric to compare the performance of tynodes. 

In figure \ref{fig:11a} and \ref{fig:12a}, the yield curves were determined on tynodes that were placed in the default orientation with the window (in the silicon substrate) and the unit cell cups facing downwards. In this orientation, the TSEs are expected to be focused to the center as shown before in figure \ref{fig:4}. The extruded ribs have the same electric potential as the center part of the hexa- and octagon, so the electric field would focus TSEs to the center of the unit cell of the next tynode. In figure \ref{fig:11b} and \ref{fig:12b}, the yield curves were determined on tynodes that were placed upside down. The unit cell cups are facing upwards in this case. In this orientation as shown figure \ref{fig:13}, the corrugated membrane is relatively flat on the transmission side and focusing effect is not to be expected. Depending on the application, one might prefer one placement over the other.  

The TSE and FSE yield curves in figure \ref{fig:11} and \ref{fig:12} are typical for tynodes. They are similar to the curves that were reported in ref. \cite{Chan2020} for flat Al\textsubscript{2}O\textsubscript{3}/TiN/Al\textsubscript{2}O\textsubscript{3} tynodes. A TSE curve has two distinct features: the threshold energy and the maximum TSE. The threshold energy $E_c$ coincides with the onset of the TSE curve and is defined as the required PE energy to generate TSEs, which depends on the thickness and the material of the film. The maximum TSE yield $\delta_T^{\text{max}}$ is achieved by using the optimal PE energy, at which the PE is relatively efficient in transferring energy near the exit surface of the film to generate TSEs. When the PE energy is further increased, the film will become electron transparant ($\eta_T\rightarrow1$), which results in a lower energy transfer and TSE yield. The FSE yield curve also has an onset, which depends on the film thickness and the required energy for a PE to penetrate the film. For higher energies, the film becomes electron transparent and the curve approaches 1. As a consequence, the amount of energy that a PE can transfer before it pass through a tynode is lower. 

The octagon pattern has a maximum transmission yield $\sigma_T^{\text{max}}$ of $2.15 (\SI{3.15}{\keV})$ and $1.85 (\SI{3.15}{\keV})$ when its cells face down- and upwards, respectively. For both measurement, the critical energy $E_c$ is $\sim\SI{1.15}{\keV}$. The critical energy is the PE energy for which the first TSEs are generated and is correlated to the membrane thickness. The difference in $\sigma_T^{\text{max}}$ for the same membrane is solely due to the 3D structure of the corrugated membrane (figure \ref{fig:13}). PEs are directional and are expected to hit the membrane perpendicularly. When the cells are facing downwards, the PEs can enter the ribs and generate TSEs from the walls and at the bottom. When the cells are facing upwards, a PE can generate TSEs at the top of the rib and on the walls, but since the angular distribution of (transmission) secondary electron emission is a cosine distribution, TSE are more likely to be recaptured. 

The hexagon honeycomb pattern has a maximum transmission yield $\sigma_T^{\text{max}}$ of $1.7 (\SI{3.05}{\keV})$ and $1.85 (\SI{3.05}{\keV})$ when placed upside down. For both measurements, the critical energy is again $\sim\SI{1.15}{\keV}$. In this case, the yield is lower when the unit cells are facing downwards. This is in contradiction with the mechanism shown in figure \ref{fig:13}. However, the reduction in transmission yield can be explained by the window size of the silicon frame, which, for the hexagonal pattern, has a width of $\SI{1.25}{\mm}$ and SEs are more likely to be recaptured by the silicon frame. In comparison, the window of the membrane with the octagon pattern has a width of \SI{4.2}{\mm}, so recapture by the silicon frame is minimal in this case. 

The maximum transmission yield of multi-layered membranes have been measured on flat membranes with similar thicknesses in ref \cite{Chan2020} and are summarized in table \ref{tab:1}. A bi-layer TiN/Al\textsubscript{2}O\textsubscript{3} film with a thickness of $5.7+\SI{25}{\nm}$ had a $\sigma_T^{\text{max}}$ of $2.1 (\SI{2.55}{\keV})$ and $E_c$ of $\SI{1}{\keV}$. A tri-layer Al\textsubscript{2}O\textsubscript{3}/TiN/Al\textsubscript{2}O\textsubscript{3} film with a thickness of $12.5+5.7+\SI{12.5}{\nm}$ had a $\sigma_T^{\text{max}}$ of $2.7 (\SI{2.75}{\keV})$ and $E_c$ of $\SI{1}{\keV}$. The tri-layer film is similar to the tri-layer film presented in this paper. However, the TiN layer is deposited by sputtering for the former, while ALD is used in the latter. The slight shift in critical energy $E_c$ of $\SI{200}{\eV}$ and the shift of the max peak by $400$ to $\SI{600}{\eV}$ indicates that the films presented in this paper are thicker in comparison. The difference in thickness can be due to the different deposition techniques of the TiN layer, but also due to residues from the process as seen in figure \ref{fig:9a} (black dots) and \ref{fig:10b} (flakes). Also, the lower $\sigma_T^{\text{max}}$ for the membranes presented in this work might be caused by the residues on the membrane surface. 

The transmission yield of a metamaterial tynode with the hexagonal pattern has also been determined in a different setup: the TyTest  \cite{VanDerReep2020}. The TyTest is a rudimentary TiPC assembled in a dedicated vacuum system, which consists of a single multiplication stage (Tynode), an e-gun as electron source and a TimePix chip as read-out. The transmission yield is in good agreement for PEs with energy of 1.2 - 1.6 keV, but has a bigger spread for PEs with energies of 1.7 and 1.8 keV. 

\begin{table}
\caption{\label{tab:1} Comparison between flat multi-layered membranes and corrugated membranes with similar thicknesses. The arrows $\downarrow,\uparrow$ indicates the direction the unit cells are facing.}  
\smallskip
\centering
\begin{tabular}{| c | c | c | c | c | c | c | c | c |}
\hline
Type & $d_{Al_2O_3} $ & $d_{\text{TiN}}$ & $d_{Al_2O_3}$ & $\sigma_{T}^\text{max}$ & $E_T^\text{max}$ & $E_c$ & Reference\\
 & (\SI{}{nm}) & (\SI{}{nm}) & (\SI{}{nm}) & & (\SI{}{keV}) & (\SI{}{keV}) & \\
\hline
Bi-layer & - & 5.7 & 25 & 2.1 & 2.55 & 0.95 & \cite{Chan2020}\\
Tri-layer & 12.5 & 5.7 & 12.5 & 2.7 & 2.75 & 0.95 & \cite{Chan2020}\\
Octagon $\downarrow$ & 10 & 5 & 15 & 2.15 & 3.15 & 1.15& this work\\
Octagon $\uparrow$ & 10 & 5 & 15 & 1.85 & 3.15 & 1.15& this work\\
Hexagon $\downarrow$ & 10 & 5 & 15 & 1.7 & 3.05 & 1.15 & this work\\
Hexagon $\uparrow$ & 10 & 5 & 15 & 1.85 & 3.05 & 1.15 & this work\\

\hline
\end{tabular}
\end{table}

\subsection{Active area \& Collection efficiency}
\label{ssec:resactive}

\begin{figure}
  \centering
	\subfloat[\label{fig:14a}]{\includegraphics[width=7.5cm]{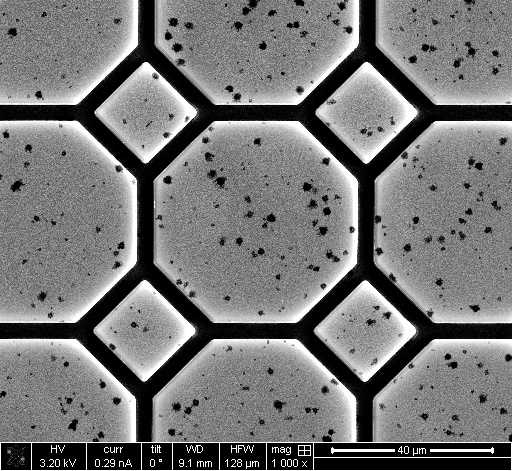}} \
	\subfloat[\label{fig:14b}]{\includegraphics[width=7.5cm]{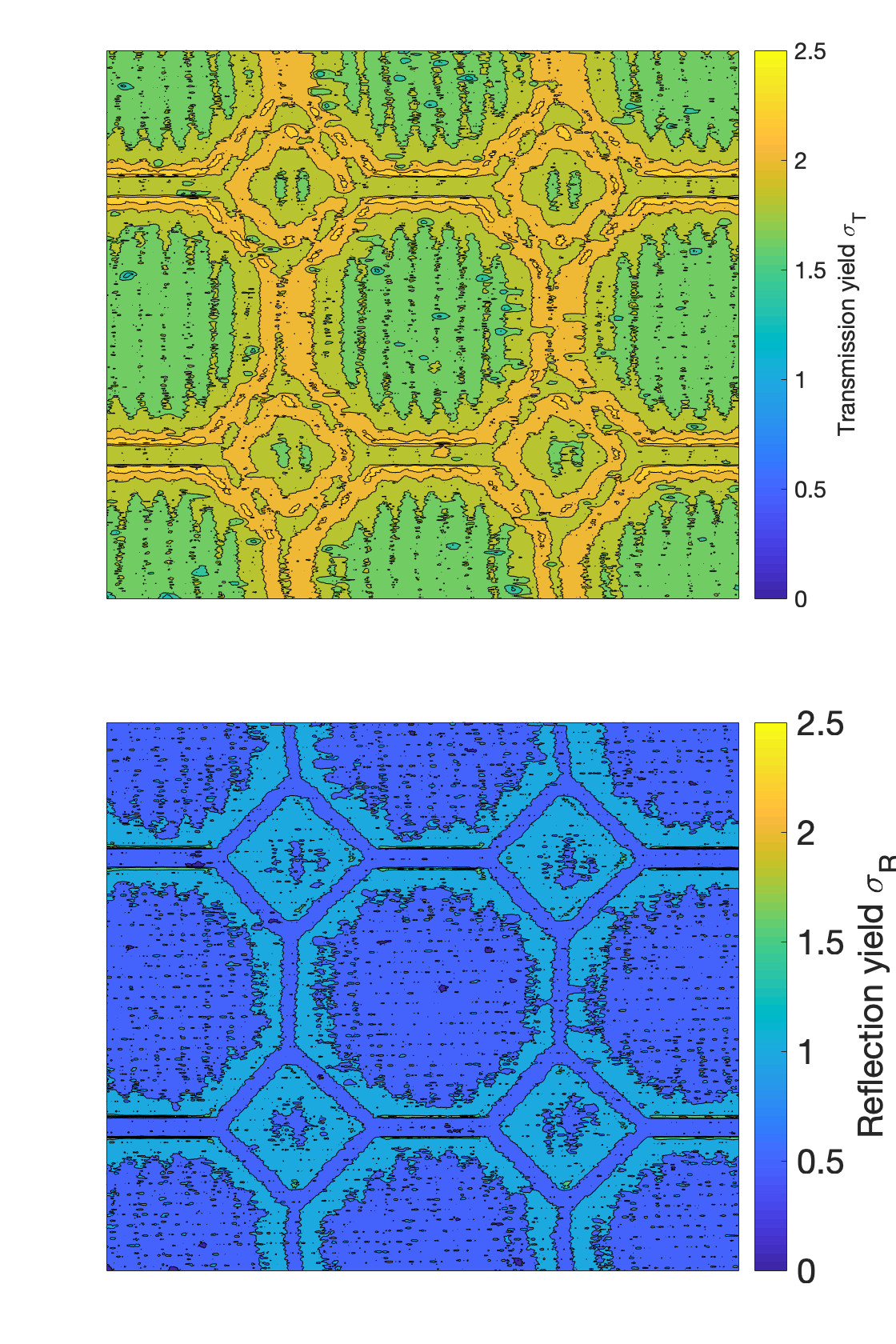}}
     \caption{Yield map of an octagon tynode with ribs facing downwards. The reflection yield map matches the SEM image. The former is determined from the sample current, while the latter is determined from SE emission from the sample. The glow near the edge of the octagon/square is visible on both pictures. The transmission yield near the vertical walls is higher due to FSEs that both generate SEs from the bottom of the cups as well from the walls.} \label{fig:14}
\end{figure}

\begin{figure}
  \centering
	\subfloat[\label{fig:15a}]{\includegraphics[width=7.5cm]{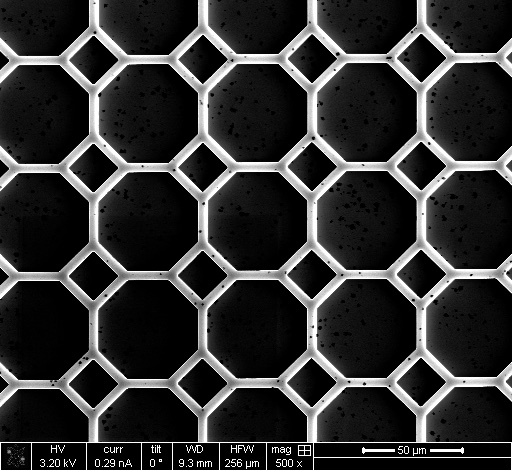}} \
	\subfloat[\label{fig:15b}]{\includegraphics[width=7.5cm]{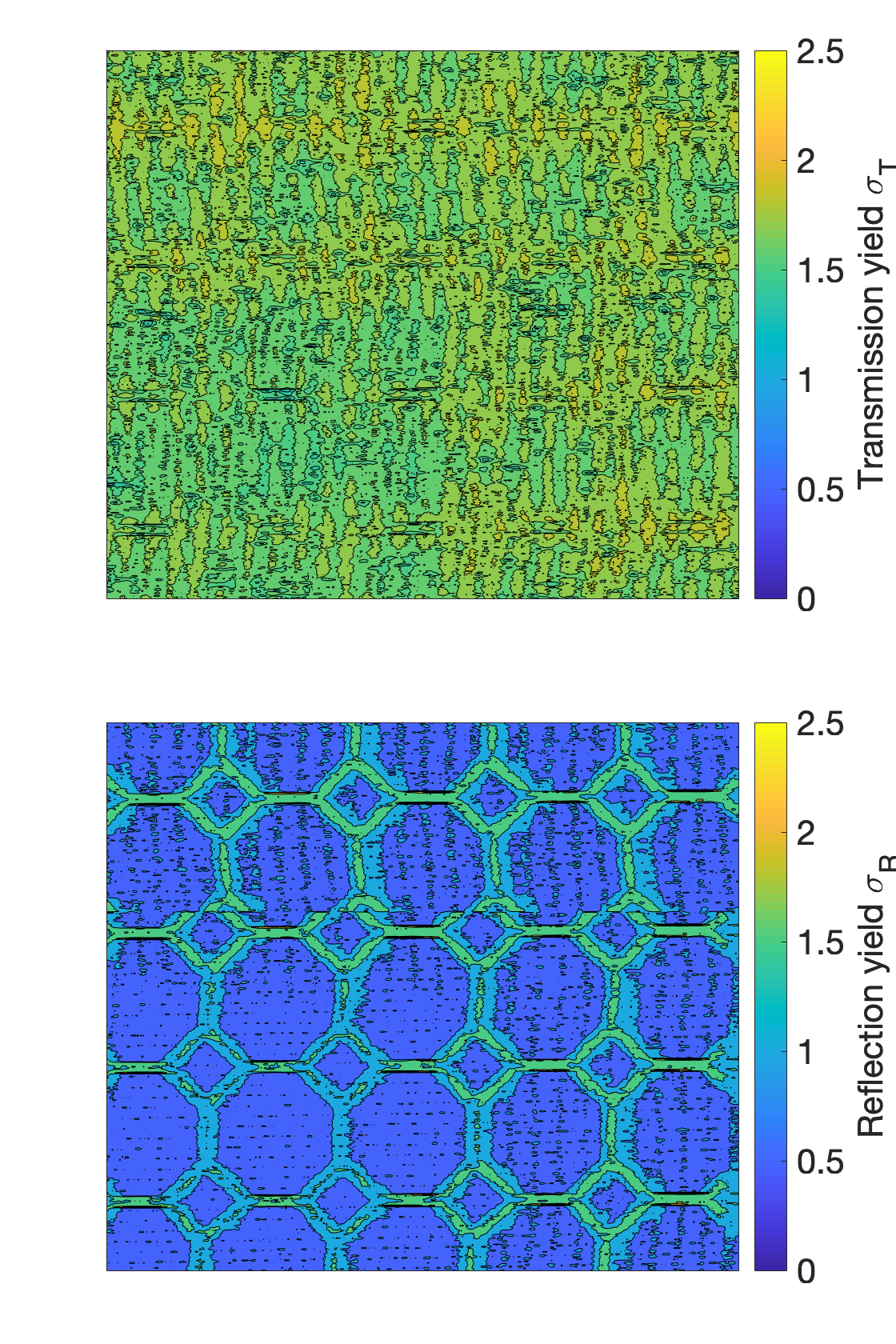}}
     \caption{Yield map of an octagon tynode with ribs facing upwards. The transmission yield map is more uniform. In reflection, the vertical walls of the ribs generate a lot more reflection SEs compared to the cups, since the probability of recapture is lower.} \label{fig:15}
\end{figure}

The active area of a tynode can be determined by measuring transmission yield as a function of the coordinates on the surface of a tynode. In figure \ref{fig:14b}, the yield map is given for a tynode with an octagon pattern. The transmission yield is higher than one on the entire surface. The active surface is thus also near 100\%. However, there is some variation in yield across the surface of the corrugated membrane. The 'hotspots' on the yield map are near the perpendicular walls of the unit cell cups. A PE transmitting through a membrane is expected to lose energy, but will still have sufficient energy to induce reflection secondary electron emission on the walls if the PE is redirected towards it. Primary electrons in bulk retain 30 to 40\% of their initial energy $E_0$ at the depth of their range $R(E_0)$ \cite{Fitting2004}. The probability that this second emission process occurs increases as the point of entry of the PE is closer to the walls of the cups.  

When the membrane is placed with the ribs facing upwards (\ref{fig:15a}), the situation is slightly different. There is no focusing effect within the unit cells and TSEs that are generated within the ribs at the top are more likely to be recaptured (figure \ref{fig:13}). The transmission yield map is more uniform in this case (figure \ref{fig:15b}), but the ribs are more pronounced in reflection as can be seen in the SEM image (figure \ref{fig:15a}) and reflection yield map. In this case, back-scattered electrons can induce a second emission process on the wall of the cups and increase the RSE yield.  

The collection efficiency of a detector is defined as the ratio of the total number of photoelectrons that are emitted from the photocathode and the number of electrons that are able to trigger a signal in the detector. The active area of the tynodes is near 100\% in both cases, which improves the collection efficiency of TiPC. Unfortunately, the hexagonal honeycomb membranes have a triangular lattice that does not match the square lattice of a pixel chip. They are therefore less suitable for TiPC. The octagonal pattern has the same lattice spacing as the pixel pads of a TimePix chip. When the octagons are aligned, the focussing effect of each tynode will bundle the electron avalanche in columns above each pixel (figure \ref{fig:4}). It can be argued that photoelectrons that are generated above the ribs or the squares in the corner of each unit cell will not be focused. They end up next to the pixel pads and will not trigger a pixel. If only the surface area of the octagon is taken, then the collection efficiency is 68\%. If we assume that the SEs generated within the ribs might contribute as well, then it is 82\%. The remaining 18\% of the surface are the squares. 

The collection efficiency can be improved by using a different corrugated pattern for the first tynode, which directs the TSEs away from the squares. Also, the collection efficiency can increase when the ribs are shaped differently by using a U- or V-shape. TSEs generated at the bottom of each rib will be forced into one of the octagonal cells in this case. This can be achieved by using different etching recipes for DRIE or using wet chemical etching (KOH) for anisotropic patterns.

\section{Conclusion \& Outlook}
\label{sec:conclusion}

We have shown that the active surface of tynodes can be increased by forming corrugated membranes, which have enhanced mechanical properties. The collection efficiency of TiPC can be improved by using these metamaterial tynodes. The corrugated pattern is formed by conformally depositing a multi-layered Al\textsubscript{2}O\textsubscript{3}/TiN/Al\textsubscript{2}O\textsubscript{3} film on a 3D silicon mold. The film is functionalized as a tynode by encapsulating a TiN layer to provide in-plane conductivity. The performance of the metamaterial tynode in terms of TEY is lower compared to flat membranes with similar composition. However, by optimizing the thickness of the ALD TiN layer in this process, the performance can be improved.

For TiPC, the TEY of the tynodes needs to be improved to 4 or higher. One way to achieve this is to consider a different secondary electron emission material, such as ALD MgO, which has a reflection secondary electron yield of 4.8 as we have reported in ref. \cite{Prodanovic2018b}. In addition, thermal annealing (up to \SI{1100}{\celsius}) and chemical treatments further improved the reflection secondary electron yield of ALD MgO to 7.2 and 6.8, respectively. We have also shown that it is feasible to fabricate ALD MgO tynodes in ref. \cite{Prodanovic2017}. A transmission SEY of 2.9 was measured on a tynode with a thickness of \SI{5}{nm}. The fabrication process presented in this work can be adapted to replace ALD Al\textsubscript{2}O\textsubscript{3} with ALD MgO. Thermal and/or chemical treatments can be considered as well.  

Improvements can be made to the design of the metamaterial tynode. The ribs can be improved by rounding the cross section into U- or V-shape, which will enhance the focusing effect: SEs that are generated within the walls of these ribs will be directed to the center of the unit cell and effectively increase the active surface. The number of pixels that the tynode covers can be increased to 256x256 by sacrificing a few rows of pixels to accommodate a silicon frame and divide the surface in smaller sections. Furthermore, the mechanical properties of the metamaterial tynode can be improved by altering the dimensions of the unit cells and or by including vertical side walls \cite{Davami2015}. For instance, the spring constant scales quadratically with the height of the plate, which is a design parameter that can be optimized. 

\acknowledgments
This work is supported by the ERC-Advanced Grant 2012 MEMBrane 320764. We would like to thank the Else Kooi Lab for providing the training, knowledge and facilities that were required to manufacture the tynodes. Many thanks to H. Akthar, W.J. Landgraaf C. Hansson, S. Tao, J. Smedley and T.v.d. Reep for their contributions to the MEMBrane project.  

\bibliographystyle{JHEP}
\bibliography{main.bbl}


\appendix

\section{Secondary electron yield map}
\label{sec:yieldmap}

A SEM image is constructed by measuring the SE emission from a specimen, while an electron beam is scanned over its surface. The speed of a scan is determined by the dwell time, the time the electron beam irradiates a surface area corresponding with a pixel in a SEM image. The line time is the time to acquire one row of the image, which is the number of pixels in the row + additional time to reposition the beam onto the front of the next row. The frame time is the time to acquire the entire image. The resolution is the number of pixels in a row times the number of pixels in a column.

The transmission yield as a function of the coordinate of a SEM image can be obtained by measuring the transmission current during image acquisition. The background current is measured before the start, while the E-beam is blanked. Image acquisition starts, while the sample, grid and collector currents are measured simultaneously. After acquisition is done, the E-beam is blanked again. The background current is measured afterwards once more. The SEM settings for the measurement are: dwell time = $\SI{1}{\ms}$, Resolution:=512 x 442, linetime = $\SI{560}{\ms}$, frametime = $\SI{4.2}{\min}$. 

The yield map reconstruction is performed with the same principle as image reconstruction in a SEM. The current is registered as a function of time. The background current is determined by the first and lasts few seconds. A measurement point as registered by the Keithley source meter can be acquired with a maximum rate of $\SI{333}{\per\s}$ or $\SI{3.3}{\ms}$ per data point. As such, each line only has 163 data points and needs to be stretched/extended into 3.3 pixels. The data points of 'missing' pixels are calculated by taking the average of two neighbouring data points. After reconstructing the lines, they are cut into $\SI{540}{\ms}$ sections and rearranged in a 2D matrix. Each line is then matched to the corresponding line in the SEM image.
  
The difficulty lies in removing artefacts that can arise due to timing delay. There can be a delay in either the SEM or the Keithley in which an extra pause is registered. The image is then distorted. These glitches can be removed manually. 

\end{document}